\documentclass[journal]{IEEEtran}

\ifCLASSINFOpdf
\else
  \usepackage[dvips]{graphicx}
\fi
\usepackage{url}

\usepackage{graphicx}
\usepackage{multirow,multicol}
\usepackage[linesnumbered,ruled,vlined]{algorithm2e}
\usepackage{amsmath}
\usepackage{setspace}
\usepackage{subfigure}
\usepackage{booktabs}
\usepackage{cellspace}
\usepackage{makecell}
\usepackage{amssymb}

\begin{document}

\title{Statistical Beamformer Exploiting Non-stationarity and Sparsity with Spatially Constrained ICA \\ for Robust Speech Recognition}

\author{Ui-Hyeop Shin and Hyung-Min Park$^*$, \IEEEmembership{Senior Member, IEEE}
\thanks{This work was partly supported by the National Research Foundation of Korea (NRF) and the Commercialization Promotion Agency for R\&D Outcomes (COMPA) grant funded by the Korea government (MSIT) (RS-2023-00237117) and Institute of Information \& communications Technology Planning \& evaluation (IITP) grant funded by the Korea government (MSIT) (No. 2022-0-00989, Development of artificial intelligence technology for multi-speaker dialog modeling).}
\thanks{The authors are with the Department of Electronic Engineering, Sogang University, Seoul 04107, Republic of Korea (e-mail: hpark@sogang.ac.kr).}}

\markboth{Journal of \LaTeX\ Class Files, Vol. 14, No. 8, August 2015}
{Shell \MakeLowercase{\textit{et al.}}: Bare Demo of IEEEtran.cls for IEEE Journals}
\maketitle

\begin{abstract}
In this paper, we present a statistical beamforming algorithm as a pre-processing step for robust automatic speech recognition (ASR). By modeling the target speech as a non-stationary Laplacian distribution, a mask-based statistical beamforming algorithm is proposed to exploit both its output and masked input variance for robust estimation of the beamformer. In addition, we also present a method for steering vector estimation (SVE) based on a noise power ratio obtained from the target and noise outputs in independent component analysis (ICA). To update the beamformer in the same ICA framework, we derive ICA with distortionless and null constraints on target speech, which yields beamformed speech at the target output and noises at the other outputs, respectively. The demixing weights for the target output result in a statistical beamformer with the weighted spatial covariance matrix (wSCM) using a weighting function characterized by a source model. To enhance the SVE, the strict null constraints imposed by the Lagrange multiplier methods are relaxed by generalized penalties with weight parameters, while the strict distortionless constraints are maintained. Furthermore, we derive an online algorithm based on an optimization technique of recursive least squares (RLS) for practical applications. Experimental results on various environments using CHiME-4 and LibriCSS datasets demonstrate the effectiveness of the presented algorithm compared to conventional beamforming and blind source extraction (BSE) based on ICA on both batch and online processing.
\end{abstract}


\IEEEpeerreviewmaketitle

\section{Introduction}

\IEEEPARstart{I}{n} order to achieve noise-robustness in automatic speech recognition (ASR), various multi-channel pre-processing methods have been adopted such as beamforming based on a steering vector~\cite{van1988beamforming, tashev2009sound} or blind source extraction/separation (BSE/BSS) that directly extracts the target speech based on independent vector analysis (IVA) (e.g.~\cite{Kim07, Ono11, Ono12, scheibler19, ikeshita20}). However, it is known that distortions and artifacts accompanied by pre-processing can cause recognition performance degradation due to mismatch in an ASR model~\cite{iwamoto22}. To address the degradation, the ASR model can be adapted to processed data or jointly optimized with pre-processing methods (e.g.~\cite{Kumatani12, Sainath17, Minhua19, Subramanian19}).
Nevertheless, in actual applications, we often encounter cases where a pre-processing algorithm should be fit to an elaborately trained large-scale ASR model for general purposes because of a reluctance to make additional tuning for a specific scenario.
In addition, beamforming methods with distortionless constraint have successfully achieved great recognition performance by enhancing target speech with minimum distortion (e.g.~\cite{van1988beamforming, tashev2009sound, cho20}). 
Also, there is evidence that beamforming methods that show better performance on a fixed ASR model generally perform better even on the model adapted to enhanced data~\cite{menne2016rwth}.
Therefore, beamforming has attracted much interest as a pre-processing method. 

In particular, a minimum-variance distortionless response (MVDR) beamformer is frequently adopted because it can effectively suppress the noises without distortion in the steered source signal by the Lagrange multiplier method for the distortionless constraint~\cite{frost1972algorithm}. For the MVDR beamformer, the spatial covariance matrix (SCM) of the noise should be estimated to effectively suppress the noises. However, because it is hard to estimate the noise SCM accurately, the minimum-power distortionless response (MPDR) beamformer can be used, which minimizes the whole input power instead~\cite{tashev2009sound}. As a result, the MPDR beamformer can replace the noise SCM with an SCM of observations, which is easily obtained from input data.
On the other hand, combined with time-frequency (t-f) masking methods~\cite{xiao2017time,heymann16_GEV,erdogan2016improved}, the noise SCM of the MVDR can be effectively estimated by using noise masks representing how much the noise is included in each t-f bin. Especially, the mask-based MVDR has shown greatly improved performance with the noise masks generated from a neural network (NN) model trained for a specific dataset~\cite{xiao2017time}.

As an alternative, a class of statistical beamforming methods, such as maximum-likelihood distortionless response (MLDR)~\cite{cho20, Cho19}, has recently been presented, showing superior performance to the MPDR. By statistical modeling of a target signal, the input SCM for the MPDR is replaced with a weighted SCM (wSCM) characterized by the weighting function from the source model. 
Such statistical beamforming methods integrate t-f segments ``considered to be more important'' with greater weights depending on how to model the target signal. In particular, the MLDR beamformer models the target speech as a complex Gaussian distribution with time-varying variances (TVVs). As a result, the weighting function of the wSCMs in the MLDR beamformer is calculated by the reciprocal value of the TVVs.
By estimating the TVVs directly from beamforming outputs, the wSCMs are successfully obtained without prior knowledge of the target signal. The TVVs can be estimated from masked inputs if target masks are available in advance~\cite{Lopez15}. This is similar to the noise SCMs of the MVDR with the noise masks in the sense that their estimates are mainly determined by the accuracy of the masks. Because the performance of the beamformer should be unavoidably limited to either the accuracy of their beamforming outputs or the masks, such a problem was recently addressed by maximum a posterior (MAP) estimation where TVVs are estimated using a prior variance obtained from NN mask~\cite{Nakatani21}. 

In this paper, we present a generalized statistical beamforming algorithm for robust ASR. Specifically, we model the target source signal as a complex Laplacian distribution with TVVs, and propose a novel statistical beamforming algorithm to exploit both its output signals and target masks. Rather than introducing a prior distribution for the TVVs to use MAP~\cite{Nakatani21}, we directly assume the sparsity of the target source to overcome the limits caused by relying on either poor initial outputs or inaccurate masks. By assuming both the sparsity and non-stationarity of source signals, the Laplacian distribution with TVVs provides a weighting function for the wSCMs that reflects both the beamforming outputs and TVVs estimated from the masked inputs.

On the other hand, an accurate steering vector should be estimated for the beamformers. Otherwise, the output may be seriously degraded due to undesirable distortion of the target speech. This steering vector estimation (SVE) with covariance subtraction method is successfully achieved by a complex Gaussian mixture model (CGMM) of multi-channel input data, assuming that t-f segments of multi-channel observations can be categorized into noise components or noisy speech components~\cite{higuchi2016robust,higuchi2017online}. Recently, an efficient SVE method was proposed by directly utilizing the wSCM of the MLDR beamformer to estimate the normalized noise SCM~\cite{cho20}.

Similarly, we introduce a method for SVE jointly updated with the beamformer in the ICA-based framework by using an output power ratio to estimate the normalized noise SCM. This is based on an auxiliary function in ICA~\cite{Ono10} with the distortionless constraint on the target output, in addition to null constraints for the noise outputs~\cite{Kim15}. The resulting beamformer is derived in the same form as MLDR, as seen in~\cite{ikeshita21}.
Compared to direct BSE without any constraints~\cite{scheibler19,ikeshita20}, such SVE followed by beamforming methods can ensure stability as pre-processing for ASR. The target SCM is obtained by subtracting a normalized noise SCM from the corresponding SCM of observations.
Therefore, the performance of the proposed SVE depends on how accurately the noises are extracted, as well as the target speech. Inspired by a geometrically constrained IVA method that imposes a power penalty for steered interference~\cite{Mitsui18,li2020geometrically}, we extend the strict constraints of the Lagrange multiplier method to hybrid constraints, using the Lagrange multiplier and the power penalty to enhance the SVE by weakening the null constraints for the noise outputs.
Furthermore, an online beamforming and SVE algorithm based on an optimization technique of recursive least squares (RLS) is derived for practical applications.

Evaluation of beamforming and SVE methods in terms of the word error rate (WER) on the CHiME-4 dataset~\cite{Vincent17} demonstrates the effectiveness of the proposed methods.
Furthermore, to assess the versatility of the proposed methods in different environments and ASR models, we conducted an utterance-wise evaluation on the LibriCSS~\cite{Chen20} dataset. We also simulated a non-stationary situation where the position of a speech source was changed to observe the effectiveness of the online algorithms. The additional experiment also demonstrates the superior results of the proposed methods as a pre-processing technique for ASR.

The remainder of this paper is organized as follows. Section \ref{Conv_beam} describes the conventional and proposed beamforming methods. In Section \ref{Proposed_SVE}, we introduce an SVE method that uses a power ratio of ICA outputs. We then derive an online beamforming algorithm based on RLS in Section \ref{RLS}. The proposed methods are evaluated through experiments in Section \ref{exp_eval}. Finally, Section \ref{Conclusion} provides a summary of our concluding remarks.

\section{Statistical Beamforming Algorithm}
\label{Conv_beam}

In real-world environments with ambient background noise, $M$ noisy speech observations at frequency bin $k$ and frame $\tau$ in the short-time Fourier transform (STFT) domain, $\mathbf{x}_k(\tau)$, can be expressed as
\begin{equation}
\mathbf{x}_k(\tau)\hspace{-0.6mm}=\hspace{-0.7mm}[X_{1k}(\tau) ,\hspace{-0.2mm} ...,\hspace{-0.2mm} X_{Mk}(\hspace{-0.1mm}\tau\hspace{-0.1mm})\hspace{-0.2mm}]^T\hspace{-1.5mm}=\hspace{-0.5mm}\mathbf{h}_k S_k(\hspace{-0.2mm}\tau\hspace{-0.2mm})\hspace{-0.3mm} +\hspace{-0.3mm} \tilde{{\mathbf{n}}}_k(\hspace{-0.2mm}\tau\hspace{-0.2mm}),\hspace{-0.9mm}
\end{equation}
where $S_k(\hspace{-.3mm}\tau)$ and $\mathbf{h}_k$ denote the corresponding t-f segment of target speech and its steering vector. $\tilde{{\mathbf{n}}}_k(\hspace{-.3mm}\tau\hspace{-.3mm})$ represents noise components in $\mathbf{x}_k(\tau)$. Conventional beamforming methods need to estimate the steering vector $\mathbf{h}_k$ using an SVE method, and then conventional beamforming methods compute a beamforming output by $Y_k(\tau)=\mathbf{w}^H_k\mathbf{x}_k(\tau)$, where $\mathbf{w}_k$ is a beamforming filter optimized under the distortionless constraint \hspace{-.2mm}of\hspace{-.2mm} $\mathbf{w}_k^{\hspace{-.2mm}H}\mathbf{h}_k\hspace{-.5mm}=\hspace{-.5mm}1$.

\subsection{MVDR Beamformer Based on Masks (Mask-MVDR)}

Because the MVDR beamformer minimizes the power of filtered noises $\sum_{\tau=1}^T|\mathbf{w}^H_k\tilde{{\mathbf{n}}}_k(\tau)|^ 2$ where $T$ denotes the number of frames, the filter is given by
\begin{eqnarray}
\mathbf{w}_k=\frac{\mathbf{V}_{N,k}^{-1}\mathbf{h}_k}{\mathbf{h}^H_k\mathbf{V}_{N,k}^{-1}\mathbf{h}_k},
\label{MVDR_weight}
\end{eqnarray}
where $\mathbf{V}_{\hspace{-0.7mm}N,k}$ is the noise SCM defined by $\mathbf{V}_{\hspace{-0.7mm}N,k} \hspace{-0mm}=\frac{1}{T}\hspace{-0mm} \sum_{\hspace{-0mm}\tau\hspace{-0mm}=\hspace{-0mm}1}^T \hspace{-0mm} \tilde{\mathbf{n}}_k(\tau\hspace{-0mm})\tilde{\mathbf{n}}_k^{H}(\hspace{-0mm}\tau\hspace{-0mm})$.
When the target mask, $\mathcal{M}_k(\tau)$, can be estimated, for example using an NN, as a power ratio of the target speech to the corresponding noisy speech at each t-f segment, $\mathbf{V}_{N,k}$ can be effectively estimated by
\begin{eqnarray}
\mathbf{V}_{N,k} = \frac1T \sum_{\tau=1}^T {(1-\mathcal{M}_k(\tau)){\mathbf{x}}_k(\tau){\mathbf{x}}^{H}_k(\tau)}.
\label{MVDR_NSCM}
\end{eqnarray}
Instead of $1/T$ in (\ref{MVDR_NSCM}), one may use $1/\sum_{\tau=1}^T (1-\mathcal{M}_k(\tau))$, which is the sum of weights $1-\mathcal{M}_k(\tau)$. Depending on the definition of the target mask, one may apply $(1-\mathcal{M}_k(\tau))^2$ instead of $1-\mathcal{M}_k(\tau)$ if the target mask is estimated as a magnitude ratio. In this paper, however, we only focus on the power ratio mask for simplicity.

\subsection{MLDR Beamformer Based on Statistical Modeling}

\newcommand*{\Scale}[2][4]{\scalebox{#1}{$#2$}}%

Instead of the MVDR beamformer that requires the masks, a class of statistical beamforming methods including MLDR was presented based on the probabilistic modeling of the target signal. The MLDR assumes that target speech follows a complex Gaussian distribution with TVVs $\lambda_k(\tau)$~\cite{Cho19}:
\begin{eqnarray}
q\left(Y_k(\tau)\right) \propto \frac{1}{\lambda_k(\tau)}\exp{\left(-\frac{|Y_k(\tau)|^2}{\lambda_k(\tau)}\right)}.
\label{MLDR_pdf}
\end{eqnarray}
Therefore, the Lagrangian function based on negative log-likelihood function of (\ref{MLDR_pdf}) with the distortionless constraint is given by
\begin{equation}
{Q}^{(Y)}_{k} =\mathbf{w}^{\hspace{-.1mm}H}_k\mathbf{V}_k{\mathbf{w}}_k + a_k^{(l)}(\mathbf{w}^{\hspace{-.2mm}H}_k\mathbf{h}_k\hspace{-.3mm}-\hspace{-.3mm}1),
\label{MLDR_cost}
\end{equation}
where $a^{\hspace{-0.2mm}(l)}_k\hspace{-0.5mm}$\hspace{-0.2mm} is a Lagrange multiplier and $\mathbf{V}_{\hspace{-0.5mm}k\hspace{-0.2mm}}$\hspace{-0.2mm} is a wSCM of
\begin{eqnarray}
\mathbf{V}_k = \frac1T \sum_{\tau=1}^T \frac{{\mathbf{x}}_k(\tau){\mathbf{x}}_k^{H}(\tau)}{\lambda_k(\tau)}.
\label{MLDR_wSCMs}
\end{eqnarray}
By minimizing (\ref{MLDR_cost}), the MLDR beamforming filter is given by~\cite{cho20,Cho19}
\begin{eqnarray}
\mathbf{w}_k=\frac{\mathbf{V}^{-1}_k\mathbf{h}_k}{\mathbf{h}^H_k\mathbf{V}^{-1}_k\mathbf{h}_k}.
\label{MLDR_weight}
\end{eqnarray}


When TVVs $\lambda_k(\tau)$ are set to a constant, the distribution becomes stationary Gaussian. Therefore, the wSCM becomes the input SCM, which corresponds to the MPDR beamforming filter~\cite{tashev2009sound}. The TVV can be directly estimated using the beamforming output as $\lambda_k(\tau) = |Y_k(\tau)|^2$ based on an iterative approach for alternative estimations of the TVV $\lambda_k(\tau)$ and beamforming weights $\mathbf{w}_k$. 
A moving average of the output powers at adjacent frames can also improve the robust estimation of the TVVs by improving the temporal continuity of source signals~\cite{Cho19}:
\begin{eqnarray}
\lambda_k(\tau) = \frac{1}{2\tau_0+1}\sum_{t=\tau-\tau_0}^{\tau+\tau_0}|Y_k(t)|^2,
\label{MLDR_TVVs}
\end{eqnarray}
where $2\tau_0+1$ is the number of adjacent frames to be averaged. Therefore, before updating alternatively $\mathbf{w}_k$ and $\lambda_k(\tau)$, the MLDR beamformer requires initialization of beamforming weights $\mathbf{w}_k$ or TVVs $\lambda_k(\tau)$.
If the target mask $\mathcal{M}_k(\tau)$ is available, the TVVs can be estimated using the masks (Mask-MLDR)~\cite{Lopez15}:
\begin{eqnarray}
\lambda_k(\tau) = \frac{1}{2\tau_0+1}\hspace{-1mm}\sum_{t=\tau-\tau_0}^{\tau+\tau_0}\hspace{-2mm}\mathcal{M}_k(t) \overline{|X_k(t)|}^2,
\label{MLDR_Mask}
\end{eqnarray}
where $\overline{|X_k(\tau)|}$ is the median value of $\{|X_{mk}(\tau)|,~m=1,\cdots\hspace{-.7mm},M\}$. Although one may apply $\mathcal{M}_k^2(\tau)$ instead of $\mathcal{M}_k(\tau)$ using a magnitude mask in (\ref{MLDR_Mask}), we only consider the power ratio mask for simplicity as in the case of (\ref{MVDR_NSCM}). This Mask-MLDR beamformer does not require iterative updates because the TVVs are uniquely estimated by (\ref{MLDR_Mask}). 

To incorporate the mask-based TVVs of (\ref{MLDR_Mask}) into the source model of target speech, a conjugate prior can be assumed as inverse-Gamma distribution $IG(\lambda_k(\hspace{-.3mm}\tau); \alpha_{\hspace{-.3mm}\lambda}, \beta_{\hspace{-.3mm}\lambda}) = \beta_{\hspace{-.3mm}\lambda}\alpha_{\hspace{-.3mm}\lambda} \Gamma\hspace{-.3mm}(\hspace{-.3mm}\alpha_{\hspace{-.3mm}\lambda}\hspace{-.3mm})^{\hspace{-.3mm}-\hspace{-.3mm}1}\hspace{-.5mm}\lambda_k(\hspace{-.3mm}\tau\hspace{-.3mm})^{\hspace{-.2mm}-(\alpha_{\hspace{-.3mm}\lambda}\hspace{-.3mm}+\hspace{-.3mm}1)}\hspace{-.5mm} \exp\hspace{-.5mm}{(\beta_{\hspace{-.3mm}\lambda}/\hspace{-.3mm}\lambda_k(\hspace{-.3mm}\tau\hspace{-.3mm})\hspace{-.3mm})}$ with $\hspace{-.3mm}\beta_{\hspace{-.3mm}\lambda\hspace{-.3mm}}$ set to TVVs from (\ref{MLDR_Mask}). The TVV can be obtained by MAP using the masked inputs as a prior (Mask-P-MLDR)~\cite{Nakatani21}: 
\begin{equation}
\label{MLDR_Mask_MAP}
  \lambda_k(\tau) = \frac{1}{2\tau_0\hspace{-0.2mm}+1}\hspace{-0.5mm}\sum_{t=\tau-\tau_0\hspace{-1.8mm}}^{\tau+\tau_0}\hspace{-0.4mm}\frac{{{|Y\hspace{-0.2mm}_k(t)|}^2\hspace{-1mm} + \hspace{-0.4mm}\mathcal{M}_k(t)\overline{|X\hspace{-0.2mm}_k(t)|}^2}}{\alpha_\lambda+2},
\end{equation}
which reflects both masked inputs and beamforming outputs. $\alpha_\lambda$ is simply set to 1.


\subsection{Generalized Statistical Beamforming}
In the MLDR beamforming, the target source model of (\ref{MLDR_pdf}) can be generalized to super-Gaussian distribution by
\begin{equation}
q\left( Y\hspace{-.5mm}_k\left( \tau \right) \right) \propto \frac{1}{\lambda_k\left( \tau \right)}\hspace{-.5mm}\exp\left(-\frac{\left| Y_k\left(\tau \right) \right|^{2\beta}\hspace{-.5mm}}{\lambda^\beta_k\left(\tau \right)} \right),
\label{source_model}
\end{equation}
where $\beta$ is a shape parameter, and $0 < \beta < 1$ is required for super-Gaussianity. Then, the wSCM in (\ref{MLDR_wSCMs}) is extended to
\begin{eqnarray}
\label{V}
\mathbf{V}_k = \frac1T\sum_{\tau=1}^T {\phi_k(\tau)}\mathbf{x}_k\left(\tau \right)\mathbf{x}^{H}_k\left(\tau \right),
\end{eqnarray}
where $\phi_k(\tau)$ is the weighting function~\cite{Ono12, Ono10}
given by
\begin{equation}
\phi_k(\tau) = \beta\frac{\left| Y_k\left(\tau \right) \right|^{2\beta-2}}{\lambda^{\beta}_k\left(\tau \right)}.
\label{general_phi}
\end{equation}
The TVV $\lambda_k\left(\tau \right)$ can be directly estimated by $\lambda_k\left(\tau \right) = \beta^{\frac{1}{\beta}}\left| Y_k\left(\tau \right)\right|^2$ in the maximum-likelihood sense. 
When $\beta=1$, the distribution is expressed as (\ref{MLDR_pdf}) and the weighting function is given by 
\begin{equation}
  \label{phi_t_MLDR}
  \phi_k(\tau)=\frac{1}{\lambda_k(\tau)},
\end{equation}
which makes (\ref{V}) equal to (\ref{MLDR_wSCMs}) of conventional MLDR beamforming. 

Strictly speaking, unlike the MPDR and MLDR, the Mask-MVDR beamformer does not have a weighting function $\phi_k(\tau)$ because its output filter is not obtained from the wSCM $\mathbf{V}_k$ based on statistic modeling but from an estimate of the noise SCM in (\ref{MVDR_NSCM}). Nevertheless, one may consider the noise SCM as a virtual wSCM by regarding the weighting function for the Mask-MVDR as $1-\mathcal{M}_k(\tau)$.

\subsection{Proposed Sparse MLDR Beamformer Using a Complex Laplacian Distribution with TVVs}

\label{prop_lap}
As a target source model, we can consider both the non-stationarity and sparsity of speech by a complex Laplacian distribution as 
\begin{eqnarray}
q\left( Y_k(\tau ) \right) \propto \frac{1}{\lambda_k\left(\tau \right)}\exp\left(- \frac{\left| Y_k\left(\tau \right) \right|}{\sqrt{\lambda_k\left(\tau \right)}}\right),
\label{ICA_pdf}
\end{eqnarray}
where the source variances are assumed to be time-varying and $\beta=1/2$. Then, the weighting function is given as
\begin{eqnarray}
\phi_k(\tau) = \frac{1}{2\sqrt{\lambda_k(\tau )}\left| Y_k(\tau) \right|}.
\label{phi_t_ICA}
\end{eqnarray}
It is noted that $\phi_k(\tau)$ in (\ref{phi_t_ICA}) contains both the TVV and the beamforming output. Although the TVV can be estimated by $\lambda_k(\tau) = |Y_k(\tau)|^2/4$, the TVV estimate directly obtained from $Y_k(\tau)$ makes no meaningful difference between ($\ref{phi_t_ICA}$) and (\ref{phi_t_MLDR}) with TVVs estimated by (\ref{MLDR_TVVs}). Instead, if the target mask is available, we can estimate the TVV with the moving average by
\begin{eqnarray}
\lambda_k(\tau) = \frac{1}{4(2\tau_0+1)}\hspace{-1mm}\sum_{t=\tau-\tau_0}^{\tau+\tau_0}\hspace{-1mm}{\mathcal{M}_k(t)\overline{|{X}_k(t)|}^2}.
\label{ICA_TVVs}
\end{eqnarray}
Then, $\phi_k(\tau)$ in (\ref{phi_t_ICA}) utilizes target speech estimates from both $Y_k(\tau)$ as the beamforming output 
and observations masked by the target masks $\mathcal{M}_k(\tau)$. We refer to this beamformer as mask-based sparse MLDR (Mask-S-MLDR). With $\beta=1/2$, the weighting function considers the beamforming outputs and masked inputs with the same exponent. 
One may obtain a better beamformer when $\beta$ is a different value between 0 and 1. However, for a distribution with TVVs, the performance difference according to the value of $\beta$ was not critical in our experience. Therefore, the Laplacian distribution in (\ref{ICA_pdf}) is considered to obtain a mathematically simple beamformer.

\vspace{0mm}
\section{Steering Vector Estimation \\ Based on ICA with the Spatial Constraints}
\label{Proposed_SVE}

\subsection{Conventional SVE Based on Covariance Subtraction}

The linearly constrained beamforming methods in Section~\ref{Conv_beam} require a steering vector. 
After modeling noisy speech observations ${\mathbf{x}}_k(\tau)$ by a CGMM, the steering vector $\mathbf{h}_k$ has been effectively estimated by the principal eigenvector of the target SCM\footnote{An estimate of $\mathbf{h}_k$ is normalized by the L2-norm $\|\mathbf{h}_k\|_2$ for stable estimation. Also, the phase of $\mathbf{h}_k$ is aligned across the frequency bins as $\mathbf{h}_k/e^{j\theta_{1k}}$ where $\theta_{1k}$ is the phase of reference (first) channel.}.
Since target speech is uncorrelated with noises, the target SCM can be estimated by the subtraction given by
\begin{eqnarray}
\mathbf{R}_{{\mathbf{s}},k} = \mathbf{R}_{{\mathbf{x}},k}- \mathbf{R}_{{\mathbf{n}},k},
\label{CGMM}
\end{eqnarray}
where $\mathbf{R}_{{\mathbf{s}},k}$, $\mathbf{R}_{{\mathbf{n}},k}$, and $\mathbf{R}_{{\mathbf{x}},k}$ denote the normalized SCMs of a target, noises, and observations at frequency bin $k$, respectively. $\mathbf{R}_{{\mathbf{x}},k}$ and $\mathbf{R}_{{\mathbf{n}},k}$ can be estimated by
\begin{eqnarray}
\label{in_SCM}
\hspace{-7mm}\mathbf{R}_{{\mathbf{x}},k}&\hspace{-2.7mm}=&\hspace{-2.7mm}\frac{1}{T}\sum_{\tau=1}^T {\mathbf{x}}_k(\tau){\mathbf{x}}^H_k(\tau), 
\\[-2pt]
\label{n_SCM}
\hspace{-7mm}\mathbf{R}_{{\mathbf{n}},k}&\hspace{-2.7mm}=&\hspace{-2.7mm}\frac{1}{\sum_{\tau=1}^T\hspace{-1mm}{r}_{\mathbf{n},k}(\tau)}\hspace{-1mm}\sum_{\tau=1}^T{r}_{\mathbf{n},k}(\tau) {\mathbf{x}}_k(\tau){\mathbf{x}}^H_k(\tau),
\end{eqnarray}
where $r_{\mathbf{n},k}(\tau)$ denotes a ratio representing noise contribution in observations, which can be estimated based on the CGMM~\cite{higuchi2016robust,higuchi2017online}. 

Replacing $r_{\mathbf{n},k}(\tau)$ with the weighting function $\phi_k(\tau)$, which has to be calculated in beamforming filter estimation, may efficiently lead to $\mathbf{R}_{\mathbf{n},k}$~\cite{cho20}: 
\begin{equation}
\mathbf{R}_{{\mathbf{n}},k}=\frac{1}{\sum_{\tau=1}^T\hspace{-1mm}{\phi}_k(\tau)}\hspace{-1mm}\sum_{\tau=1}^T{\phi}_k(\tau) {\mathbf{x}}_k(\tau){\mathbf{x}}^H_k(\tau).
\label{wSCMS}
\end{equation}
Because the weighting function $\phi_k(\tau)$ is, for example, calculated as a reciprocal value of the TVV in the Gaussian source model, it emphasize t-f components with small value of TVVs $\lambda_k(\tau)$ which correspond to noise-dominant segments, and vice versa~\cite{cho20, Cho19}. Therefore, the wSCM $\mathbf{V}_k$ effectively estimates the noise SCM by accumulating the contributions of noise.

The $r_{\mathbf{n},k}(\tau)$ can be estimated by $r_{\mathbf{n},k}(\tau)=1-\mathcal{M}_k(\tau)$ if the target mask is available. However, to improve the SVE further, the target mask can be effectively incorporated with SVE methods based on covariance subtraction (CGMM, wSCM, and proposed methods explained later) by using masked observations $\hspace{-1mm}\sqrt{\hspace{-.2mm}\mathcal{M}_k(\hspace{-.2mm}\tau)}\mathbf{x}_k(\hspace{-.2mm}\tau)$ instead of $\mathbf{x}_k(\hspace{-.2mm}\tau)$ to suppress corrupted t-f segments in (\ref{in_SCM}), (\ref{n_SCM}), and (\ref{wSCMS}), like~\cite{cho20, Cho19}. 

\subsection{BSE of Target Speech without SVE Based on IVA}
On the other hand, the formulation for the beamforming problem can be extended to model the spatial mixing process of noises as well as target speech in BSE formulation by defining mixing matrix $\mathbf{A}_k$. The $M$ noisy speech observations, $\mathbf{x}_k(\tau)$, can be re-modeled as
\begin{eqnarray}
\mathbf{x}_k(\tau) = \mathbf{A}_k\begin{bmatrix}
S_k(\tau)\\
\mathbf{{n}}_k(\tau)
\end{bmatrix},
\label{ICA_formulation}
\end{eqnarray}
where the mixing matrix $\mathbf{A}_k$ includes $\mathbf{h}_k$ as the first column and $\mathbf{n}_k(\tau)$ denotes a source vector of $\tilde{{\mathbf{n}}}_k(\tau)$. The mixing environment is assumed to be determined for simplicity. 
Then, let us consider the demixing model:
\begin{eqnarray}
\begin{bmatrix}
Y_k(\tau)\\ \mathbf{z}_k(\tau)
\end{bmatrix} =\mathbf{W}_k\mathbf{x}_k(\tau),
\label{BSS_form}
\end{eqnarray}
where $\mathbf{W}\hspace{-.3mm}_k\hspace{-1mm}=\hspace{-1mm}[\mathbf{w}_1\hspace{-.2mm}_k, ...,\hspace{-.5mm}\mathbf{w}_{\hspace{-.3mm}M}\hspace{-.3mm}_k]^H\hspace{-.5mm}$ denotes a demixing matrix.
Without loss of generality, let us assume that $Y\hspace{-.2mm}_k(\hspace{-.3mm}\tau)\hspace{-.7mm} =\hspace{-.7mm} S_k(\hspace{-.3mm}\tau)$ whereas $\mathbf{z}_k(\hspace{-.0mm}\tau)$ represents noise outputs. Then, assuming different source models for target speech and noises, the likelihood function for the target sources can be extended to include the one for noise sources as
\begin{eqnarray}
Q_{\hspace{-0mm}k} \hspace{-0mm}= \hspace{0mm}\frac12{\mathbf{w}_{1k}^{H}\mathbf{V}_{\hspace{-0.1mm}k}\mathbf{w}_{1k}} \hspace{-0mm} + \frac12\hspace{-0mm} \hspace{-0mm}\sum_{m=2}^M\hspace{-0mm}{\mathbf{w}_{\hspace{-0mm}mk}^{H}\hspace{-0.1mm}\mathbf{V}\hspace{-0mm}_{\mathbf{z},k}\mathbf{w}\hspace{-0mm}_{mk}} \hspace{10mm}\nonumber \\[-0pt] -\hspace{-0mm}\log{\hspace{-0mm}|\hspace{-0mm}\det{\hspace{-0mm}\mathbf{W}_{\hspace{-0mm}k}}|},
\label{IVA_cost}   
\end{eqnarray}
which leads to the auxiliary function of ICA~\cite{Ono10}.
$\mathbf{V}_{\hspace{-.5mm}\mathbf{z},k}$ is the wSCM for noises obtained by replacing $\phi_k(\hspace{-.3mm}\tau)$ with $\phi_{\mathbf{z},k}(\tau)$ in (\ref{V}).

In the conventional BSE, target speech is directly extracted by optimizing (\ref{IVA_cost}) with a source modeled as a multivariate Gaussian distribution 
in the IVA~\cite{Ono12} as
\begin{equation}
q( Y_1(\tau),...,Y_K(\tau) ) \propto \hspace{-.4mm}\frac{1}{\tilde{\lambda}^K(\tau)}\exp\hspace{-.5mm}\left(\hspace{-1mm}-\frac{\sum_{k=1}^K|Y_k(\tau)|^2}{\tilde{\lambda}( \tau)}\right)
\label{IVA_model}
\end{equation}
to address the random permutation problem without the explicit spatial constraints. $K$ is the number of frequency bins, and $\tilde{\lambda}(\tau)$ is a shared TVV along frequency bins which is estimated by $\tilde{\lambda}(\tau)=\frac1K\sum_{k=1}^K|Y_k(\tau)|^2$. 
Then, the weighting function is given as $\phi_k(\tau) = 1/\tilde{\lambda}(\tau)$~\cite{Ono12}. 
Noises are often assumed to follow a stationary Gaussian distribution (e.g.~\cite{scheibler19,ikeshita20,Nakatani21,ikeshita21}).
Then, the weighting function for the noises is given by $\phi_{\mathbf{z},k}(\tau) = 1$. Such a distinct modeling enforces the target speech to be extracted at a desired channel.


\subsection{Proposed SVE Using Output Powers of ICA with the Spatial Constraints}
\label{Prop_SVE}

Estimating $\mathbf{R}_{\mathbf{n},k}$ based on $\phi_k(\tau)$ might be unstable because its value range is so wide that $\mathbf{R}_{\mathbf{n},k}$ can excessively depend on some large values of the weighting function $\phi_k(\tau)$. Although the conventional SVE methods are effective, the ratio $r_{\mathbf{n},k}(\tau)$ can be more elaborately estimated by utilizing noise outputs as well. In other words, the ratio $r_{\mathbf{n},k}(\tau)$ can be calculated using the ICA outputs from (\ref{BSS_form}). After $\mathbf{W}_k$ is updated, $Y_k(\tau)$ and $\mathbf{z}_k(\tau)$ obtained by (\ref{BSS_form}) should be normalized to estimate an accurate power ratio by the minimal distortion principle~\cite{matsuoka2002minimal}:
\begin{eqnarray}
\text{diag}\left(\mathbf{W}^{-1}_k\right)\begin{bmatrix}
Y_k(\tau)\\ \mathbf{z}_k(\tau)
\end{bmatrix}=\begin{bmatrix}
\hat{S}_k(\tau)\\
\hat{\mathbf{n}}_k(\tau)
\end{bmatrix},
\label{MDP}
\end{eqnarray}
where $\text{diag}(\cdot)$ returns the diagonal matrix with off-diagonal component being zeros. Using the normalized ICA outputs, the ratio can be estimated by
\begin{eqnarray}
r_{\mathbf{n},k}(\tau) = \frac{\| \mathbf{\hat{n}}_k(\tau) \|_2^2}{|\hat{S}_k(\tau)|^2+ \| \mathbf{\hat{n}}_k(\tau) \|_2^2 },
\end{eqnarray}
where $\|\cdot\|_2$ denotes the L2-norm of a vector.

Especially, the auxiliary function of (\ref{IVA_cost}) can be augmented by the distortionless and null constraints to extract and cancel the target speech at the target and noise outputs regarding the steering vector $\mathbf{h}_k$.
In (\ref{ICA_formulation}), when the mixing matrix $\mathbf{A}_k \hspace{-.9mm}=\hspace{-.5mm} [\mathbf{h}_k|\mathbf{D}_k]$ where $\mathbf{D}_k$ is an $M\hspace{-.9mm}\times\hspace{-.8mm}(M\hspace{-0.4mm}-\hspace{-0.4mm}1)$ matrix to get $\tilde{{\mathbf{n}}}_k(\tau) = \mathbf{D}_k \mathbf{n}_k(\tau)$, $\mathbf{x}_k(\tau)$ is expressed as 
$\mathbf{x}_k(\tau) = \mathbf{h}_kS_k(\tau) + \mathbf{D}_k{{\mathbf{n}}}_k(\tau).$
Because $\mathbf{A}\hspace{-.1mm}_k\mathbf{e}_1\hspace{-.3mm}=\hspace{-.3mm}\mathbf{h}_k$, we can obtain $Y_k(\tau)=S_k(\tau)$ similar to the conventional beamforming by introducing the distortionless constraint of $\mathbf{w}_{1k}^H\mathbf{h}_k=1$ in the BSE formulation of (\ref{BSS_form}). In addition, the null constraints for the noises are given as $\mathbf{w}_{mk}^H\hspace{-.3mm}\mathbf{h}_k=0, \ 2 \le m \le M$~\cite{Kim15}.
Accordingly, instead of initializing the demixing matrix $\mathbf{W}_{\hspace{-0.2mm}k}$ to the identity matrix as in conventional ICA and IVA for BSS, we can initialize it by
\begin{equation}
\label{W_init}
  \mathbf{W}_k = \left[\mathbf{h}_k | \mathbf{e}_2, ..., \mathbf{e}_M \right]^{-1},
\end{equation}
which corresponds to the identity matrix replacing the $\mathbf{e}_1$ with the steering vector $\mathbf{h}_k$. We can initialize $\mathbf{h}_k$ to all-one vector $\mathbf{1}$, assuming that the target speaker is perpendicular to a microphone array, or to the $\mathbf{e}_1$, which results in initializing $\mathbf{W}_k$ to the identity matrix. 

Therefore, similarly to (\ref{MLDR_cost}), the auxiliary function of (\ref{IVA_cost}) can be extended to Lagrangian function with the Lagrange multipliers $a_{mk}^{(l)}$ for the constraints (ICA-LC):
\begin{equation}
Q_k^{(l)}\hspace{-.7mm}= Q_k + a_{1k}^{(l)}(\mathbf{w}_{1k}\mathbf{h}_k \hspace{-.5mm}-\hspace{-.5mm} 1) + \hspace{-.7mm}\sum_{m=2}^{M}\hspace{-.5mm}a_{mk}^{(l)}\mathbf{w}_{mk}\mathbf{h}_k.
\label{Aux_L}
\end{equation}
Then, optimizing the Lagrangian function using the constraints $a_{1k}^{(l)}$ with respect to $\mathbf{w}_{1k}$ gives (See Appendix \ref{appendix:ICA_LC}.)
\begin{eqnarray}
\mathbf{w}_{1k}=\frac{\mathbf{V}^{-1}_k\mathbf{h}_k}{\mathbf{h}^H_k\mathbf{V}^{-1}_k\mathbf{h}_k},
\label{gen_beamform}
\end{eqnarray}
which has the same form as (\ref{MLDR_weight}). The statistical beamforming filter of (\ref{gen_beamform}) based on the ICA with distortionless and null constraints is reduced to the same form of MLDR beamforming, which is similarly derived in \cite{ikeshita21}. The noise outputs can be obtained by updating $\mathbf{w}_{mk}, 2\hspace{-.2mm}\leq\hspace{-.2mm}m\hspace{-.2mm}\leq\hspace{-1mm}M$ from (\ref{Aux_L}) and its derivation can be found in Appendix \ref{appendix:ICA_LC}. Consequently, we can calculate the ratio $r_{\mathbf{n},k}(\tau)$ from the beamforming output as well as noise outputs to cancel the target speech, which enables the joint updates of the beamforming weights and the steering vector in the same ICA framework.

On the other hand, in multi-dimensional ICA~\cite{Cardoso98}, noise sources were modeled by a multidimensional subspace expressed as $q\left( \mathbf{z}_k\left(\tau \right) \right)\propto \exp\left(-\left\|\mathbf{z}_k(\tau)\right\|_2\right)$,
which is a spherical multivariate Laplacian distribution. 
Its weighting function is derived as
\begin{eqnarray}
\phi_{\mathbf{z},k}(\tau) = \frac{1}{2\|\mathbf{z}_k(\tau)\|_2}.
\label{phi_n_MICA}
\end{eqnarray}
The weighting function modeled by multi-dimensional ICA may deal with the various diffuse noises by reflecting the sparsity of the noises. 

\subsection{A Proposed Hybrid Approach to Impose the Spatial Constraints}
\label{HC_sec}
With incorrect steering vector estimates before convergence, the strict spatial constraints by Lagrange multipliers may prevent finding the correct demixing matrix $\mathbf{W}_k$. Instead, one can consider more mild constraints.
In \cite{Mitsui18,li2020geometrically}, spatial constraints (corresponding to the distortionless constraint for target speech and null constraints for noises) by the steering vector $\mathbf{h}_k$ were imposed for better separation, which can be formulated by the auxiliary function augmented with the power penalty terms of the spatial constraints (ICA-PC): 
\begin{equation}
\label{cost:ICA-PC}
Q_{k}^{(p)}\hspace{-.5mm}= Q_k+{a}_{1k}^{\hspace{-.1mm}(p)}|\mathbf{w}^{\hspace{-.1mm}H}_{1k}\mathbf{h}_k-1|^2 \hspace{-.2mm}
+\hspace{-.2mm}{a}_{\mathbf{z},k}^{\hspace{-.1mm}(p)}\hspace{-.1mm}\sum_{m=2}^M\hspace{-.2mm}|\mathbf{w}^{H}_{mk}\mathbf{h}_k|^2\hspace{-.1mm},
\end{equation}
where ${a}_{1k}^{(p)}$ and ${a}_{\mathbf{z},k}^{(p)}$ are parameters determining weights for the distortionless and null constraints for target speech and noises, respectively.

However, a slight distortion in estimated speech may result in considerable recognition performance degradation in robust ASR, as mentioned above. 
Consequently, we propose a hybrid approach for beamforming by imposing the strict distortionless constraint based on a Lagrange multiplier and more flexible null constraints based on power penalties in the auxiliary function (ICA-HC):
\begin{equation}
Q_{k}^{(h)}\hspace{-1.2mm}= Q_k + a_{1k}^{\hspace{-0.2mm}(l)}(\mathbf{w}_{1k}^H\mathbf{h}_k-1)+{a}^{\hspace{-0.2mm}(p)}_{\mathbf{z},k}\hspace{-.4mm}\sum_{m=2}^M\hspace{-.3mm}\left|\mathbf{w}_{mk}^H\hspace{-0.2mm}\mathbf{h}_k\right|^2\hspace{-1.5mm}.
\label{cost_prop}
\end{equation}
Through a derivation and some approximation, 
the target weights $\mathbf{w}_{1k}$ can be still updated by using (\ref{gen_beamform}) in ICA-LC. On the other hand, updates of the noise weights $\mathbf{w}_{mk},~2\leq m \leq M,$ are exactly the same as in ICA-PC:
\begin{eqnarray}
\hspace{-10mm}\mathbf{H}_{\mathbf{z},k}\hspace{-2.5mm} &=& \hspace{-2mm}\mathbf{V}_{\mathbf{z},k}+{a}_{\mathbf{z},k}^{(p)}\hspace{0mm}\mathbf{h}_k\mathbf{h}^H_k,
\label{Noise_penalty_0}
\\[4pt]
\hspace{-10mm}\tilde{\mathbf{w}}_{mk} \hspace{-2.5mm} &=& \hspace{-2mm}\left(\mathbf{W}_k\mathbf{H}_{\mathbf{z},k} \right)^{-1}\mathbf{e}_m,
\label{Noise_penalty_1}
\\[3pt]
\hspace{-10mm}{\mathbf{w}}_{mk}\hspace{-2.5mm}&=&\hspace{-2mm}{\tilde{\mathbf{w}}_{mk}}\hspace{.5mm}/{\sqrt{\tilde{\mathbf{w}}_{mk}^H\mathbf{H}_{\mathbf{z},k}\tilde{\mathbf{w}}_{mk}}},~2 \leq m \leq M.
\label{Noise_penalty_3}
\end{eqnarray}



\section{Derivation of an RLS-based Online \\ Algorithm for Beamforming and SVE}
\label{RLS}

We also present a frame-by-frame online RLS algorithm for ICA-based beamforming and SVE using the formulation of $Y_k(t;t\hspace{-.5mm}-\hspace{-.5mm}1) = \mathbf{w}^H_{1k}(t\hspace{-.3mm}-\hspace{-.3mm}1)\mathbf{x}_k(t)$ with the target weights updated at ($t-1$)-th frame.

\subsection{An RLS-Based Online MLDR Beamformer}

From the cost for the batch method of (\ref{MLDR_cost}), the RLS-based MLDR cost at the $t$-th frame is expressed with forgetting factor $\alpha$ as
\begin{equation}
\tilde{Q}^{(\hspace{-.2mm}Y\hspace{-.3mm})}_{k}\hspace{-.5mm}(t) \hspace{-.5mm}=\hspace{-.5mm} {\mathbf{w}_k^{H}(t)\hspace{-.1mm}\mathbf{V}_{\hspace{-0.4mm}k}\hspace{-0.2mm}(\hspace{-.1mm}t)\mathbf{w}_k(t)} + a_k^{\hspace{-.3mm}(l)}\hspace{-.3mm}(\hspace{-.1mm}t)(\mathbf{w}_k^{\hspace{-.4mm}H}\hspace{-.3mm}(\hspace{-.1mm}t)\mathbf{h}_k(\hspace{-.1mm}t) - 1)\hspace{-.3mm},
\end{equation}
where the wSCM at the $t$-th frame is calculated by
\begin{eqnarray}
\mathbf{V}_k(t) = \frac{1}{\sum_{\tau=1}^t \hspace{-0.7mm}\alpha^{t-\tau}}\sum_{\tau=1}^t \alpha^{t-\tau}\phi_k(\tau){{\mathbf{x}}_k(\tau){\mathbf{x}}^{H}_k(\tau)}.
\end{eqnarray}
Therefore, $\mathbf{V}_k(t)$ can be recursively obtained as
\begin{equation}
\mathbf{V}_{\hspace{-0.2mm}k}(\hspace{-0.2mm}t) \hspace{-0.2mm}=\hspace{-0.2mm} \rho(t)\mathbf{V}_{\hspace{-0.2mm}k}(\hspace{-0.2mm}t\hspace{-0.2mm}-\hspace{-0.2mm}1) + (\hspace{-0.2mm}1\hspace{-0.2mm}-\hspace{-0.2mm}\rho(t))\phi_k(\hspace{-0.1mm}t)\mathbf{x}_k(\hspace{-0.1mm}t)\mathbf{x}_k^{H}\hspace{-0.2mm}(\hspace{-0.1mm}t).
\label{Online_V}
\end{equation}
where $\rho(t) = 1-{1}/{{\sum_{\tau=1}^{t}\alpha^{t-\tau}}}$. The inverse matrix $\mathbf{U}\hspace{-.4mm}_k\hspace{-0.2mm}(t)=\mathbf{V}_k^{\hspace{-0.2mm}-\hspace{-.2mm}1}\hspace{-0.4mm}(t)$ can be directly updated by using the matrix inversion lemma~\cite{woodbury50}:
\begin{eqnarray}
\mathbf{U}_k\hspace{-0.1mm}(\hspace{-0.1mm}t ) \hspace{-0.4mm}=\hspace{-0.3mm} \frac1{\rho(t)}\mathbf{U}_k\hspace{-0.1mm} (\hspace{-0.3mm}t\hspace{-0.7mm}-\hspace{-0.7mm}1) \hspace{47mm} \nonumber \\[1pt] \hspace{2.5mm} - \frac{\mathbf{U}_k\hspace{-0.1mm}(\hspace{-0.3mm}t\hspace{-0.7mm}-\hspace{-0.7mm}1)\mathbf{x}_k(\hspace{-0.3mm}t)\mathbf{x}_k^H(t)\mathbf{U}\hspace{-0.3mm}^H_k(\hspace{-0.1mm}t\hspace{-0.7mm}-\hspace{-0.7mm}1)}{\rho^2\hspace{-.3mm}(\hspace{-.1mm}t\hspace{-.1mm})\hspace{-.3mm}/\hspace{-.3mm}(\hspace{-.5mm}(1\hspace{-0.8mm}-\hspace{-0.8mm}\rho(t)\hspace{-.5mm})\phi_{\hspace{-0.1mm}k}(\hspace{-0.2mm}t)\hspace{-.5mm})\hspace{-.6mm}+\hspace{-.6mm}\rho(t)\mathbf{x}_k^H\hspace{-0.7mm}(\hspace{-0.1mm}t)\mathbf{U}_{\hspace{-0.3mm}k}\hspace{-0.1mm}(\hspace{-0.2mm}t\hspace{-0.8mm}-\hspace{-0.8mm}1\hspace{-.1mm})\mathbf{x}_k\hspace{-0.1mm}(\hspace{-0.1mm}t)}\hspace{-.5mm}.\hspace{-1.8mm}
\label{Online_UY}
\end{eqnarray}
Then, the update of the beamforming filter at the $t$-th frame can be obtained by
\begin{eqnarray}
\label{online_MLDR}
\mathbf{w}_k(t)=\frac{\mathbf{U}_k(t)\mathbf{h}_k(t)}{\mathbf{h}^H_k(t)\mathbf{U}_k(t)\mathbf{h}_k(t)}.
\end{eqnarray}

Also, if the weighting function $\phi_k(t)=1/\lambda_k(t)$ using the Gaussian distribution of (\ref{MLDR_pdf}), $\lambda_k(t)$ can be recursively estimated by
\begin{eqnarray}
\lambda_k(t) = \gamma\lambda_k(t-1) + (1-\gamma)|Y_k(t;t-1)|^2,
\label{online_TVV}
\end{eqnarray}
where $\gamma$ is a smoothing factor. Note that we use $Y\hspace{-.7mm}_k(t;t\hspace{-.5mm}-\hspace{-.5mm}1)$ obtained by the previous beamforming filter $\mathbf{w}_k(t\hspace{-.5mm}-\hspace{-.5mm}1)$ because $Y_k(t;t)=\mathbf{w}^H_k(t)\mathbf{x}_k(t)$ is not available. If the target mask is available, it can be estimated by replacing $|Y\hspace{-.6mm}_k(t;t\hspace{-0.5mm}-\hspace{-.5mm}1)|^2$ with $\mathcal{M}_k(t)\overline{|{X}_k(t)|}^2$ in (\ref{online_TVV}), similar to (\ref{MLDR_Mask}) in batch processing. Or, based on MAP, we can update $\lambda_k\hspace{-0.2mm}(t)$ by using both the values as
\begin{eqnarray}
\lambda_k(t) = \gamma\lambda_k(t-1) + \nonumber \hspace{45mm} \\ (1-\gamma)
\frac{\mathcal{M}_k(t)\overline{|{X}_k(t)|}^2+|Y_k(t;t-1)|^2}{\alpha_\lambda + 2},
\label{MLDR_P_online}
\end{eqnarray}
which corresponds to (\ref{MLDR_Mask_MAP}) in Mask-P-MLDR.
For the Mask-S-MLDR beamformer using Laplacian distribution of (\ref{ICA_pdf}), the weighting function at the t-th frame is calculated as
\begin{eqnarray}
\phi_k(t) = \frac{1}{2\sqrt{\lambda_k( t )}\left| Y_k( t;t-1) \right|}.
\label{Online_phi}
\end{eqnarray}
From (\ref{ICA_TVVs}), the TVVs of the Mask-S-MLDR beamformer at the $t$-th frame can be easily updated by
\begin{eqnarray}
\lambda_k(t) = \gamma\lambda_k(t-1)+\frac{1\hspace{-.3mm}-\hspace{-.3mm}\gamma}{4}\mathcal{M}_k(t)\overline{|{X}_k(t)|}^2.
\label{Prop_Online_TVV}
\end{eqnarray}

\subsection{An RLS-Based Online Updates for SVE with ICA-HC}
\label{ICA_RLS}
Unlike the conventional online IVA algorithm~\cite{Taniguchi14} using auto-regressive estimation of the wSCM, we derive an RLS-based online beamformer that may result in a generalized online beamformer. The auxiliary function at the $t$-th frame is defined as
\begin{eqnarray}
\tilde{Q}_k\hspace{-0.4mm}(t) \hspace{-0.7mm}=\hspace{-0.7mm} {\mathbf{w}_{\hspace{-0.4mm}1k}^{H}\hspace{-0.4mm}(t)\mathbf{V}_{\hspace{-0.6mm}k}\hspace{-0.2mm}(t)\mathbf{w}_{\hspace{-0.3mm}1k}(t)}\hspace{-.5mm}+\hspace{-1.5mm}\sum_{m=2}^M\hspace{-0.9mm}{\mathbf{w}_{\hspace{-0.4mm}mk}^{H}(t) \mathbf{V}\hspace{-0.7mm}_{\mathbf{z},k}(t)\mathbf{w}\hspace{-0.3mm}_{mk}(t)} \hspace{2mm} \nonumber \\[0pt] -\log{\hspace{-0.2mm}|\hspace{-0.2mm}\det\hspace{-0.4mm}{\mathbf{W}_{\hspace{-0.5mm}k}(\hspace{-0.2mm}t)}|}\hspace{-0.2mm},\hspace{-1.2mm}
\label{RLS_cost}
\end{eqnarray}
Also, the update equations of $\mathbf{V}_{\mathbf{z},k}(t)$ and $\mathbf{U}_{\mathbf{z},k}(t) =\mathbf{V}_{\mathbf{z},k}^{-1}(t)$ can be obtained by replacing $\phi_k(\tau)$ with $\phi_{\mathbf{z},k}(\tau)$ in (\ref{Online_V}) and (\ref{Online_UY}), respectively.
For the online updates of the wSCMs, the weighting functions of (\ref{phi_n_MICA}) can be computed by
\begin{eqnarray}
\phi_{\mathbf{z},k}(t) = \frac{1}{2\|\mathbf{z}_k(t;t-1)\|_2}.
\label{Online_phi_z}
\end{eqnarray}
For the online updates of the weights in ICA-HC, the augmented auxiliary function is given as
\begin{eqnarray}
\tilde{Q}_{k}^{(h)}(t)=\tilde{Q}_{\hspace{-.2mm}k}\hspace{-.3mm}(\hspace{-.1mm}t\hspace{-.1mm}) + a_{\hspace{-.2mm}1k}^{\hspace{-.1mm}(\hspace{-.1mm}l\hspace{-.1mm})}(\hspace{-.1mm}t\hspace{-.1mm})(\hspace{-.1mm}\mathbf{w}_{\hspace{-.1mm}1k}^H\hspace{-.2mm}(\hspace{-.1mm}t\hspace{-.1mm})\mathbf{h}_{\hspace{-.2mm}k}(\hspace{-.1mm}t\hspace{-.1mm})-1\hspace{-.2mm}) \hspace{15mm} \nonumber \\ +\ {a}_{\mathbf{z},\hspace{-.2mm}k}^{(p)}\sum_{m=2}^M|\mathbf{w}^H_{\hspace{-.1mm}mk}\hspace{-.1mm}(\hspace{-.1mm}t\hspace{-.1mm})\hspace{-.1mm}\mathbf{h}_k\hspace{-.3mm}(\hspace{-.1mm}t\hspace{-.1mm})\hspace{-.1mm}|^2\hspace{-.1mm}.
\end{eqnarray}
Similar to Subsection~\ref{HC_sec}, the update rules of the noise weights can be easily induced as follows: 
\begin{eqnarray}
\label{Online_prop_beam}
\label{Online_PC_n0_0}
\hspace{-7mm}\mathbf{H}_{\mathbf{z},k}\hspace{-.1mm}(t)\hspace{-3mm}&=&\hspace{-3mm}\mathbf{V}_{\hspace{-.1mm}\mathbf{z},k}(t)\hspace{-.1mm}+{a}_{\mathbf{z},k}^{(p)}\hspace{-0mm}\mathbf{h}_k(t)\mathbf{h}^H_k(t),
\\[5pt]
\label{Online_PC_n0_1}
\hspace{-7mm}\mathbf{G}\hspace{-0.2mm}_{\mathbf{z},k}^{(p)}\hspace{-0.2mm}(\hspace{-0.2mm}t) \hspace{-3mm}&=&\hspace{-3mm}\mathbf{U}_{\mathbf{\hspace{-0.3mm}z},k}\hspace{-0.1mm}(\hspace{-0.2mm}t)\hspace{-0.45mm}-\hspace{-0.45mm} \frac{\mathbf{U}_{\mathbf{\hspace{-0.3mm}z},k}(\hspace{-0.2mm}t)\mathbf{h}_k(\hspace{-0.2mm}t)\mathbf{h}_k^{\hspace{-0.3mm}H}(\hspace{-0.2mm}t)\mathbf{U}_{\mathbf{\hspace{-0.3mm}z},k}(\hspace{-0.1mm}t)}{1\hspace{-0.2mm}/\hspace{-0.1mm}{a}_{\mathbf{z},k}^{(p)}\hspace{-0.4mm}+\hspace{-0.4mm}\mathbf{h}_k^{\hspace{-0.3mm}H}(\hspace{-0.1mm}t)\mathbf{U}_{\mathbf{\hspace{-0.3mm}z},k}(\hspace{-0.1mm}t)\mathbf{h}_k(\hspace{-0.1mm}t) }\hspace{-0.3mm},
\\[2pt]
\label{Online_PC_n1}
\hspace{-7mm}\tilde{\mathbf{w}}_{mk}(t)\hspace{-3mm}&=&\hspace{-3mm}\mathbf{G}_{\mathbf{z},k}^{(p)}(t)\mathbf{A}_k(t)\mathbf{e}_m,
\\[6pt]
\label{Online_PC_n3}
\hspace{-7mm}{\mathbf{w}}_{mk}(t)\hspace{-3mm}&=&\hspace{-3mm}{\tilde{\mathbf{w}}_{\hspace{-.2mm}mk}(t)}\hspace{-.3mm}/\hspace{-.8mm}{\sqrt{\tilde{\mathbf{w}}_{\hspace{-.2mm}mk}^H\hspace{-.4mm}(t)\mathbf{H}_{\mathbf{z},k}\hspace{-.4mm}(t)\tilde{\mathbf{w}}_{\hspace{-.2mm}mk}\hspace{-.4mm}(t)}},\hspace{.5mm}2\hspace{-.3mm}\leq\hspace{-.3mm}m\hspace{-.3mm}\leq\hspace{-.3mm}M,
\end{eqnarray}
where $\mathbf{G}_{\mathbf{z},k}^{(p)}(t)\hspace{-.4mm}=\hspace{-.4mm}\mathbf{H}_{\mathbf{z},k}^{-1}\hspace{-.1mm}(t)$. 
Because the mixing matrix $\mathbf{A}_k(t)\hspace{-.4mm}=\hspace{-.4mm}\mathbf{W}^{-1}_k(t)$ is required at every update of $\mathbf{w}_{mk}(t)$, it should be updated by~\cite{Taniguchi14}
\begin{eqnarray}
\mathbf{A}_k(t) = \mathbf{A}_k(t) - \frac{\mathbf{A}_k(t)\mathbf{e}_m\Delta \mathbf{w}^H_{mk}(t)\mathbf{A}_k(t)}{{1+\Delta \mathbf{w}^H_{mk}(t)\mathbf{A}_k(t)\mathbf{e}_m}},
\label{Online_A}
\end{eqnarray}
where $\Delta\mathbf{w}_{mk}(t) = \mathbf{w}_{mk}(t)-\mathbf{w}_{mk}(t-1)$.

The SVE can be performed at every frame using online updates of the normalized SCM of observations and normalized noise SCMs, given by
\begin{eqnarray}
\mathbf{R}_{{\mathbf{x}},k}(t)=\frac{1}{\sum_{\tau=1}^t \alpha^{t-\tau}}\sum_{\tau=1}^t \alpha^{t-\tau} {\mathbf{x}}_k(\tau){\mathbf{x}}^H_k(\tau),
\end{eqnarray}
\begin{equation}
\mathbf{R}_{\hspace{-0.2mm}{\mathbf{n}}\hspace{-0.2mm},k}\hspace{-0.3mm}(t) \hspace{-0.3mm}=\hspace{-0.4mm}\frac{1}{\sum_{\hspace{-.3mm}\tau=1}^t \hspace{-0.6mm} \alpha^{\hspace{-0.1mm}t\hspace{-0.2mm}-\hspace{-0.2mm}\tau}{r}_{\hspace{-0.3mm}\mathbf{n},k}(\hspace{-0.1mm}\tau)}\hspace{-0.6mm}\sum_{\tau=1}^t \hspace{-0.6mm} \alpha^{\hspace{-0.3mm}t\hspace{-0.2mm}-\hspace{-0.2mm}\tau}\hspace{-.4mm}{r}_{\hspace{-0.2mm}\mathbf{n},k}(\hspace{-0.1mm}\tau) {\mathbf{x}}_k\hspace{-0.2mm}(\hspace{-0.1mm}\tau\hspace{-0.2mm}){\mathbf{x}}^{\hspace{-0.2mm}H}_k\hspace{-0.3mm}(\hspace{-0.3mm}\tau)\hspace{-.3mm}.\hspace{-1mm}
\end{equation}
Then, $\mathbf{R}_{{\mathbf{x}},k}(t)$ and $\mathbf{R}_{{\mathbf{n}},k}(t)\hspace{-0.3mm}$ are recursively updated by
\begin{eqnarray}
\hspace{-8mm}\mathbf{R}_{{\mathbf{x}},k}(t)&\hspace{-3mm}=\hspace{-2.8mm}&\rho(t)\mathbf{R}_{{\mathbf{x}},k}(t-1) +(1\hspace{-.5mm}-\hspace{-.5mm}\rho(t)){\mathbf{x}}_k(t){\mathbf{x}}\hspace{-.5mm}^H_k(t),
\label{Online_SCM}
\\[7pt]
\hspace{-6.5mm}\mathbf{R}_{{\mathbf{n}},k}(t)&\hspace{-3mm}=\hspace{-2.8mm}&\tilde{\rho}_k{(t)}\mathbf{R}_{{\mathbf{n},k}\hspace{-.3mm}}(\hspace{-.2mm}t\hspace{-.7mm}-\hspace{-.7mm}1) \hspace{-.7mm}+\hspace{-.7mm}(\hspace{-.3mm}1\hspace{-.7mm}-\hspace{-.7mm}\tilde{\rho}_k(t)\hspace{-.4mm}){\mathbf{x}}_k(\hspace{-.2mm}t){\mathbf{x}}^H_k(t)\hspace{-.2mm},
\label{Online_NSCM}
\end{eqnarray}
where
\vspace{2mm}
\begin{equation}
\tilde{\rho}_k(t) = 1 - \frac{r_{\mathbf{n},k}(t)}{\sum_{\tau=1}^t \alpha^{t-\tau}{r}_{\mathbf{n},k}(\tau)}.
\end{equation}
The output power ratio is calculated by
\begin{eqnarray}
r_{\mathbf{n},k}(t) = \frac{\| \mathbf{\hat{n}}_k(t) \|_2^2}{|\hat{S}_k(t)|^2+ \| \mathbf{\hat{n}}_k(t) \|_2^2 },
\label{power_ratio}
\end{eqnarray}
where the outputs normalized by the minimal distortion principle~\cite{matsuoka2002minimal} are obtained by
\begin{eqnarray}
\text{diag}\left(\mathbf{A}_k(t-1)\right)\begin{bmatrix}
Y_k(t;t-1)\\ \mathbf{z}_k(t;t-1)
\end{bmatrix}=\begin{bmatrix}
\hat{S}_k(t)\\
\hat{\mathbf{n}}_k(t)
\end{bmatrix}.
\label{Online_MDP}
\end{eqnarray}
Also, we empirically found that the proposed SVE methods based on ICA can be unstable especially when the position of a target source are changed because the proposed SVE requires the noise sources to be extracted in online manner, too. Therefore, robust estimation of $r_{\hspace{-.5mm}\mathbf{n},k}(t)$ can be improved as
\begin{eqnarray}
r_{\mathbf{n},k}(t) = \frac{P_{\mathbf{n},k}(t)}{|\hat{S}_k(t)|^2+ P_{\mathbf{n},k}(t) },
\label{power_ratio_recur}
\end{eqnarray}
by replacing $\| \mathbf{\hat{n}}_k(t) \|_2^2$ in (\ref{power_ratio}) with $P_{\mathbf{n},k}(t)$ recursively updated by $P_{\mathbf{n},k}(t)\hspace{-.3mm}=\hspace{-.3mm}\gamma_\mathbf{n}P_{\mathbf{n},k}\hspace{-.1mm}(t\hspace{-.2mm}-\hspace{-.2mm}1)\hspace{-.1mm}+\hspace{-.1mm}(1\hspace{-.2mm}-\hspace{-.2mm}\gamma_\mathbf{n})\|\hspace{-.2mm}\mathbf{\hat{n}}_k(t)\hspace{-.2mm}\|_2^2$ with smoothing factor $\gamma_\mathbf{n}$ assuming the stationarity of the noise sources. 
After the noise SCM is obtained, the target SCM is estimated by 
\begin{equation}
\mathbf{R}_{{\mathbf{s}},k}(t) = \mathbf{R}_{{\mathbf{x}},k}(t)- \nu\mathbf{R}_{{\mathbf{n}},k}(t),
\label{CS_online}
\end{equation}
where $\nu$ is a scale factor for stable online processing~\cite{cho20}. The scale factor $\nu$ is useful for the online SVE because subtracting an inaccurate estimate of the noise SCM 
may result in a disastrous SVE.
The value of $\nu$ can be set to even zero to enable stable SVE, even if there is little noise in the initial frames where the estimate of the noise SCM is likely to be inaccurate. 
Finally, the steering vector is obtained by finding the principal eigenvector of $\mathbf{R}_{{\mathbf{s}},k}(t)$ with the normalization.
\begin{algorithm}[t]
\label{Total_online_algorithm}
\setstretch{1.1}
\small
 \caption{Proposed online Mask-S-MLDR beamforming and SVE algorithm based on ICA-HC} 
\SetAlgoLined
Initialize $\mathbf{W}_k$, $\mathbf{A}_k$, $\alpha$, $\gamma$, $\gamma_\mathbf{n}$, $\nu$, and ${a}_{\mathbf{z},k}^{(p)}$
\;
\For{$t = 1,...,T$}
{
\For{$k = 1,...,K$}
{
  Compute outputs $\begin{bmatrix} \hspace{-0.1mm}Y\hspace{-0.3mm}_k(\hspace{-0.3mm}t;t\hspace{-0.6mm}-\hspace{-0.7mm}1\hspace{-0.1mm})\hspace{-0.9mm}\\ \hspace{-0.1mm}\mathbf{z}_k(\hspace{-0.3mm}t;t\hspace{-0.6mm}-\hspace{-0.7mm}1)\hspace{-0.9mm} \end{bmatrix} \hspace{-1.1mm}=\hspace{-0.7mm}\mathbf{W}_{\hspace{-0.3mm}k}(t\hspace{-0.5mm}-\hspace{-0.5mm}1)\mathbf{x}_k(\hspace{-0.4mm}t)$. \\
  \textbf{/* Estimate Steering Vector */}\\
  Normalize the outputs by (\ref{Online_MDP}).\\
  Update $P_{\mathbf{n},k}(t)=\gamma_\mathbf{n}P_{\mathbf{n},k}(t\hspace{-.6mm}-\hspace{-.6mm}1) + (1-\gamma_\mathbf{n})\| \mathbf{\hat{n}}_k(t) \|_2^2$.\\
  Calculate $r_{\hspace{-0.2mm}\mathbf{n},\hspace{-0mm}k}\hspace{-0.3mm}(t)$ by (\ref{power_ratio_recur}). \\ 
  Update $\mathbf{R}_{\mathbf{x},k}(\hspace{-0.3mm}t) $ and $\mathbf{R}_{\mathbf{n},k}(\hspace{-0.3mm}t) $ using (\ref{Online_SCM}) and (\ref{Online_NSCM}).\\
  Calculate $\mathbf{R}_{\mathbf{s},k}(t)$ by (\ref{CS_online}) and update $\mathbf{h}_k(t)$.\\
  \textbf{/* Update the wSCMs */}\\
  Update $\lambda_k(t)$ by (\ref{Prop_Online_TVV}).\\
  Calculate $\phi_k(t)$ and $\phi_{\mathbf{z},k}(t)$ using (\ref{Online_phi}) and (\ref{Online_phi_z}). \\
  Update $\mathbf{U}\hspace{-.2mm}_k(\hspace{-.2mm}t) $ using (\ref{Online_UY}).\\
  Update $\hspace{-.5mm}\mathbf{V}_{\mathbf{\hspace{-.8mm}z\hspace{-.2mm}},k}(\hspace{-.5mm}t)\hspace{-.5mm}$ and $\mathbf{U}_{\hspace{-.2mm}\mathbf{z},k}(\hspace{-.2mm}t) $ with (\ref{Online_V}) and (\ref{Online_UY}) after replacing $\hspace{-.5mm}\phi_k(\hspace{-.3mm}t\hspace{-.2mm})\hspace{-.5mm}$ with $\phi_{\mathbf{z},k}(\hspace{-.2mm}t\hspace{-.2mm})$.\\
  \textbf{/* Update Beamforming Output Weights */} \\
  Set $\mathbf{A}_k(t)=\mathbf{A}_k(t-1)$.\\
  Update $\mathbf{w}_{1k}(t)$ by (\ref{Online_prop_beam}). \\
  Update $\mathbf{A}_k(t)$ by (\ref{Online_A}). \\
  \textbf{/* Update Noise Output Weights */} \\
  Update $\mathbf{H}_{\mathbf{z},k}(t)$ and $\mathbf{G}^{(p)}_{\mathbf{z},k}(t)$ by (\ref{Online_PC_n0_0}) and (\ref{Online_PC_n0_1}). \\
  \For{$m=2,...,M$}
  {
    Update $\mathbf{w}_{mk}(t)$ by (\ref{Online_PC_n0_0})-(\ref{Online_PC_n3})\\
    Update $\mathbf{A}_k(t)$ by (\ref{Online_A}). \\
  }
  \textbf{/* Calculate beamforming Output */} \\
  Compute $Y_k(t;t)=\mathbf{w}_{1k}^H(t)\mathbf{x}_k(t) $. \\
}
}
\end{algorithm}
The overall RLS-based online beamforming and SVE algorithm based on ICA with the constraints is summarized in Algorithm \ref{Total_online_algorithm}.

\section{Experimental Evaluation}
\label{exp_eval}
The presented algorithms were evaluated through ASR experiments on the CHiME-4 challenge dataset~\cite{Vincent17}. The dataset was recorded using six microphones in four noisy environments, and the evaluation was based on the WERs.
The ASR system was constructed by the Kaldi toolkit~\cite{povey2011kaldi}, which was the same as in~\cite{Cho19}. The 13-order Mel-frequency cepstral coefficients feature is used for training the acoustic model, extracted from input noisy training data, corresponding close-talk microphone data for real-recorded data, and clean data simulated at the six microphones.
We also utilized the Kaldi recurrent-neural-network-based language model. For a more detailed description of the ASR system construction, please refer to~\cite{Cho19}. When analyzing using the STFT, a Hanning window with a length of 1024 samples and a shift of 256 samples was commonly applied to the data with a sampling rate of 16 kHz. For target masks, we used the ones estimated by the same NN model as in~\cite{heymann16_GEV}.

Furthermore, when obtaining the median value $\overline{|X_k(\tau)|}$ of microphone observations $\mathbf{x}_k(\tau)$, the observations at the second microphone were excluded to obtain a better estimation of TVVs. For robust processing, it is necessary to clip the weighting functions $\phi_k(\tau)$ by a value of $\phi_0$, as $\phi_k(\tau) \gets \min{(\phi_k(\tau), \phi_0)}$~\cite{Ono12, Cho19}, when calculating the wSCMs in all the statistical beamformers. The clipping value $\phi_0$ was experimentally set to $10^6$. In ICA, the weighting function $\phi_{\mathbf{z},k}(\tau)$ was also clipped by the same value for the proposed noise source model.

\subsection{Comparison of the Proposed Beamforming and SVE with Conventional Enhancement Methods}
\label{SE_vs_PropBF}

\begin{table}
\footnotesize
\caption{WERs (\%) on the CHiME-4 dataset for the baseline without any processing for input data acquired at the fifth microphone (similar to~\cite{Cho19,Barker15}) and enhancement by batch OIVA, GEV, MPDR, MVDR, and a family of MLDR beamforming methods including the proposed Mask-S-MLDR beamformers. SVE methods based on CGMM, wSCMs, and proposed ICA-HC were performed by observations masked by all-one and NN masks. SVE based on NN alone with Mask-MVDR was also compared.
}
\renewcommand{\tabcolsep}{4.6pt}
\def\arraystretch{1.08}
\begin{center}
\footnotesize
\begin{tabular}{c|c|c|cccc|c}
\hline \hline
\multirow{1}{*}{\hspace{-1.8mm}\textbf{Enhancement}\hspace{-1.8mm}}&\multirow{2}{*}{\hspace{-2mm}\textbf{Mask}\hspace{-2mm}} &\multirow{2}{*}{\hspace{-0.5mm}\textbf{SVE}\hspace{-0.5mm}}&\multicolumn{2}{c}{\textbf{ Dev. } }&\multicolumn{2}{c|}{\textbf{ Test } } & \multirow{2}{*}{\textbf{\hspace{-1.3mm}Avg.\hspace{-1.3mm}}}\\
\textbf{Method}& & &{\hspace{-0.5mm}simu.\hspace{-0.5mm}} & {\hspace{-0mm}real\hspace{0mm}} & {\hspace{-0.5mm}simu.\hspace{-0.5mm}} & {\hspace{-0mm}real\hspace{-0mm}} & \\
\hline
\hspace{-2mm}No Processing\hspace{-2mm} & - & - & 5.51 & 6.52 & 6.71 & \hspace{-1.2mm}11.63 & 7.59\\ \hline
OIVA & - & - & 4.57 & 7.11 & 6.90 & \hspace{-1.2mm}16.03 & 8.65\\
MPDR   & - & \text{\hspace{-1.4mm}CGMM\hspace{-1.4mm}} & 2.94 & {\bf 3.17} & 4.78 & 5.63 & 4.13\\
MLDR & - & \hspace{-1mm}\multirow{1}{*}{wSCM}\hspace{-1mm} & {\bf 2.77} & {3.24} & {4.21} & {4.85} & {3.76} \\
MLDR & - & {ICA-HC} & {2.79} & {3.31} & {\bf 3.77} & {\bf 4.83} & {\bf 3.68} \\

\hline
Mask-GEV & NN & - & {\bf 2.35} & 2.72 & {\bf 2.98} & 4.41 & {\bf 3.12}\\
\hspace{-2mm}Mask-P-OIVA\hspace{.3mm}\hspace{-2mm} & NN & - & 2.47 & 3.93 & 3.24 & 6.49 & 4.03\\
\hspace{-0mm}Mask-MVDR & NN & NN & {2.49} & {\bf 2.71} & 3.26 & {\bf 4.24} & 3.18\\
\hline
\hspace{-0mm}Mask-MVDR & NN & {\hspace{-1.4mm}NN\hspace{.4mm}+\hspace{.4mm}CGMM\hspace{-1.4mm}} & {2.42} & {2.51} & 3.22 & {3.89} & 3.01\\
\text{\hspace{-.5mm}Mask-MLDR\hspace{-.5mm}} & NN & \hspace{-1mm}{NN\hspace{.4mm}+\hspace{.4mm}CGMM}\hspace{-1mm} & {2.32} & {2.52} & {3.20} & {3.65} & {2.92} \\
\text{\hspace{-.5mm}Mask-P-MLDR\hspace{-.5mm}} & NN & \hspace{-1mm}{NN\hspace{.4mm}+\hspace{.4mm}CGMM}\hspace{-1mm} & {2.37} & {2.48} & {\bf 2.95} & {3.60} & {2.85} \\
{Mask-S-MLDR} & NN & \text{\hspace{-1.4mm}{NN\hspace{.4mm}+\hspace{.4mm}CGMM}\hspace{-1.4mm}} & {\bf 2.36} & {\bf 2.42} & {3.00} & {\bf 3.59} & {\bf 2.84} \\
\hline
\text{\hspace{-.5mm}Mask-MLDR\hspace{-.5mm}} & NN & \hspace{-1mm}{NN\hspace{.4mm}+\hspace{.4mm}wSCM}\hspace{-1mm} & {2.32} & {2.53} & {3.21} & {3.84} & {2.98} \\
\text{\hspace{-.5mm}Mask-P-MLDR\hspace{-.5mm}} & NN & \hspace{-1mm}{NN\hspace{.4mm}+\hspace{.4mm}wSCM}\hspace{-1mm} & {2.37} & {2.50} & {\bf 3.12} & {\bf 3.72} & {\bf 2.93} \\
{Mask-S-MLDR} & NN & \text{\hspace{-1.4mm}{NN\hspace{.4mm}+\hspace{.4mm}wSCM}\hspace{-1.4mm}} & {\bf 2.29} & {\bf 2.50} & {3.19} & {3.84} & {2.96} \\
\hline
\text{\hspace{-.5mm}Mask-MLDR\hspace{-.5mm}} & NN & \hspace{-1mm}{NN\hspace{.4mm}+\hspace{.4mm}ICA-HC}\hspace{-1mm} & {2.32} & {2.46} & {3.08} & {3.51} & {2.84} \\
\text{\hspace{-.5mm}Mask-P-MLDR\hspace{-.5mm}} & NN & \hspace{-1mm}{NN\hspace{.4mm}+\hspace{.4mm}ICA-HC}\hspace{-1mm} & {2.37} & {2.54} & {3.07} & {3.54} & {2.88} \\
{Mask-S-MLDR} & NN & \hspace{-.5mm}\text{{NN\hspace{.4mm}+\hspace{.4mm}ICA-HC}}\hspace{-.5mm} & {\bf 2.29} & {\bf 2.43} & {\bf 3.05} & {\bf 3.44} & {\bf 2.80} \\

\hline \hline
\end{tabular}
\end{center}
\label{tab:WER0}
\end{table}

In Table~\ref{tab:WER0}, various conventional methods for speech enhancement were compared with the proposed beamforming and SVE methods based on batch processing.
As a BSE method without explicit spatial constraints, over-determined IVA (OIVA)~\cite{ikeshita20} was also compared. When the masks were not used, the Gaussian source model with the shared TVVs $\tilde{\lambda}(\tau)$ was used for target speech by (\ref{IVA_model}) to solve the permutation problem. Additionally, the output was normalized using the minimal distortion principle of (\ref{MDP}).
If target masks were used, two enhancement methods without constraints were compared: mask-based generalized eigenvalue beamforming (Mask-GEV)~\cite{heymann16_GEV} and OIVA with masked inputs $\mathcal{M}_k(\tau)\overline{|X_k(\tau)|}^2$ as the prior (Mask-P-OIVA). TVVs were estimated by (\ref{MLDR_Mask_MAP}) using MAP. Note that the output $Y_k(\tau)$ of Mask-P-OIVA has scale ambiguity without spatial constraints, so $Y_k(\tau)$ was replaced with $\hat{S}_k(\tau)$ by (\ref{MDP}) when estimating the TVVs in (\ref{MLDR_Mask_MAP}).

On the other hand, two conventional beamforming and SVE methods were compared. One method was the Mask-MVDR beamformer with the CGMM-based SVE~\cite{higuchi2016robust} whose overall SVE procedure was conducted before the beamforming. Without the masks, the Mask-MVDR beamforming was replaced with MPDR beamforming. The other method was the MLDR beamformer with wSCM-based SVE~\cite{cho20} based on (\ref{wSCMS}) where the SVE and MLDR beamforming were alternately updated. 
In the MLDR beamformer with masks, the Mask-MLDR and Mask-P-MLDR beamformers were used, where TVVs were estimated by (\ref{MLDR_Mask}) directly from the masked input and by (\ref{MLDR_Mask_MAP}) using the masked input as the prior. As a proposed beamforming and SVE method, the Mask-S-MLDR with SVE based on the ICA-HC using TVVs of (\ref{ICA_TVVs}) and the stationary Laplacian noise model of (\ref{phi_n_MICA}) was evaluated. For the comprehensive study of beamforming and SVE methods, we also experimented all the other possible combination of beamforming (Mask-MLDR, Mask-P-MLDR, and Mask-S-MLDR) and SVE (CGMM, wSCM, and ICA-HC) methods when the target masks is given.
For the family of MLDR beamformers (MLDR, Mask-P-MLDR, and Mask-S-MLDR) that require iterations, $\mathbf{w}_k$ is initialized to $\mathbf{e}_1$. This means that the initial $Y_k(\tau)$ is equal to $X_{1k}(\tau)$ in batch processing. For the ICA-HC-based SVE, $\mathbf{W}_k$ is initialized using (\ref{W_init}) with $\mathbf{h}_k=\mathbf{1}$. Then, the initial $\mathbf{w}_{1k}$ for the beamformer is still equal to $\mathbf{e}_1$, while the weights for the noises are initialized to $\mathbf{w}_{mk} = \mathbf{e}_m - \mathbf{e}_1$. 

For all iterative methods, including CGMM and MLDR beamformers, the number of iterations in batch processing was set to 10. For all BSE and beamforming methods with the TVVs, $\tau_0$ was set to 1.
For the proposed SVE based on ICA-HC, the penalty weight ${a}_{\mathbf{z},k}^{(p)}$ for the null constraint was set to 1. In OIVA and Mask-P-OIVA, they were empirically initialized in the same way using (\ref{W_init}). 
For all SVE methods, the input $\hspace{-1mm}\sqrt{\hspace{-.2mm}\mathcal{M}_k(\hspace{-.2mm}\tau)}\mathbf{x}_k(\hspace{-.2mm}\tau)$ was used to estimate SCMs in (\ref{in_SCM}), (\ref{n_SCM}), and (\ref{wSCMS}) to further suppress noise components if masks were available. In particular, the posterior probability of the CGMM was fixed at a noise mask $1-\mathcal{M}_k$ for the first five iterations to sufficiently exploit the information of NN masks and further enhance performance. 
In addition, to see the effectiveness of the SVE methods (CGMM, wSCM, and ICA-HC) with the NN masks, we experimented the SVE based on the NN mask alone by setting $r_{\mathbf{n},k}(\tau)=1-\mathcal{M}_k(\tau)$ estimated from the input $\mathbf{x}_k(\tau)$ instead of masked input in the (\ref{in_SCM}), (\ref{n_SCM}), and (\ref{wSCMS}).

In Table~\ref{tab:WER0}, SVE-based methods improved performance without masks. While OIVA and GEV with masks improved recognition performance, especially for simulated data, linearly constrained beamforming methods (MPDR, MVDR, MLDR, and Mask-S-MLDR) with SVE based on (\ref{CGMM}) achieved robustness even for real-recorded data, demonstrating the stability of SVE by (\ref{CGMM}). Among the two SVE methods with masks based on Mask-MVDR beamformer, combining CGMM with NN masks showed more improved performance than simply using NN masks alone. When CGMM-based SVE was performed, MLDR-based beamforming methods generally showed better results than Mask-MVDR. 
Additionally, wSCM-based methods with a class of MLDR beamformers outperformed Mask-MVDR beamformer with CGMM-based SVE. However, although the wSCM-based SVE is computationally efficient, it is observed that it showed worse performances compared to CGMM and ICA-HC for the same beamforming methods. Furthermore, among the three SVE methods, the proposed SVE based on ICA-HC generally showed the best results with NN masks. 

Without NN masks, MLDR beamformer with proposed SVE based on ICA-HC still yielded the best results on average. Among the family of mask-based MLDR beamformers, Mask-S-MLDR generally showed the lowest averaged WER, demonstrating the effectiveness of the proposed beamformer. 
Especially, it is noted that Mask-S-MLDR provided comparable or even better performance than Mask-P-MLDR because the proposed Mask-S-MLDR effectively utilized beamforming outputs and masked inputs in the weighting function.
Moreover, proposed SVE based on ICA-HC with Mask-S-MLDR beamformer achieved better recognition performance on average than all compared enhancement methods for NN masks. These results confirm the effectiveness of both the proposed Mask-S-MLDR beamformer and SVE based on ICA-HC.

\subsection{Comparison of Proposed Online Beamforming and SVE with Conventional Online Enhancement Methods}

\begin{table}
\footnotesize

\caption{WERs (\%) on the CHiME-4 dataset for the baseline without any processing for input data acquired at the fifth microphone and enhancement by online OIVA, MLDR, and proposed Mask-S-MLDR beamformers. SVE methods based on wSCMs and proposed ICA-HC were performed by observations masked by all-one and NN masks. SVE based on NN alone with Mask-MVDR was also compared.
}

\renewcommand{\tabcolsep}{4.75pt}
\def\arraystretch{1.08}

\begin{center}
\footnotesize
\begin{tabular}{c|c|c|cccc|c}
\hline \hline
\multirow{1}{*}{\hspace{-1.5mm}\textbf{Enhancement}\hspace{-1.5mm}}&\multirow{2}{*}{\hspace{-0.8mm}\textbf{Mask}\hspace{-0.8mm}}&\multirow{2}{*}{\hspace{-0.5mm}\textbf{SVE}\hspace{-0.5mm}}&\multicolumn{2}{c}{\textbf{ Dev. } }&\multicolumn{2}{c|}{\textbf{ Test } } & \multirow{2}{*}{\textbf{\hspace{-1.3mm}Avg.\hspace{-1.3mm}}}\\
\textbf{method}& & & {\hspace{-0.7mm}simu.\hspace{-0.7mm}} & {\hspace{-0mm}real\hspace{0mm}} & {\hspace{-0.7mm}simu.\hspace{-0.7mm}} & {\hspace{-0mm}real\hspace{-0mm}} & \\
\hline
\text{\hspace{-1.5mm}No Processing\hspace{-1.5mm}} & - & - & 5.51 & 6.52 & 6.71 & \hspace{-1.2mm}11.63 & 7.59\\ \hline

OIVA &-& - & 9.64 & \hspace{-1.2mm}11.83 & \hspace{-1.2mm}10.48 & \hspace{-1.2mm}30.07 & \hspace{-1.2mm}15.51\\
MLDR &-& \text{\hspace{-1mm}wSCM\hspace{-1mm}}& {3.14} & 3.41 & {5.07} & 5.75 & 4.34\\
MLDR &-&{ICA-HC}& {\bf 3.13} & {\bf 3.31} & {\bf4.99} & {\bf 5.55} & {\bf 4.25} \\

\hline
\text{\hspace{-0.8mm}Mask-P-OIVA\hspace{-0.8mm}} & NN & - & 2.68 & 3.99 & 3.35 & 5.96 & 4.00 \\
\text{\hspace{-1mm}Mask-MVDR\hspace{-1mm}} &NN&{NN}&  3.23&  3.40&  4.31&  6.66&  4.40\\ 
\text{\hspace{-1mm}Mask-P-MLDR\hspace{-1mm}} &NN&{NN\hspace{.3mm}+\hspace{.5mm}wSCM}& 2.72 & 2.91 & {3.26} & 4.39 & 3.32\\ 
{Mask-S-MLDR} &NN&{NN+\hspace{.5mm}wSCM}& {2.67} & {2.84} & {3.18} & {4.40} & {3.27} \\
{Mask-S-MLDR} &NN&{\text{\hspace{-1mm}{NN+\hspace{.5mm}ICA-HC}\hspace{-1mm}}}& {\bf 2.61} & {\bf 2.79} & {\bf 3.17} & {\bf 4.23} & {\bf 3.20} \\

\hline \hline
\end{tabular}
\end{center}
\label{tab:On_WER}
\end{table}

\begin{table}
\footnotesize
\caption{Hyper-parameters for the proposed online Mask-S-MLDR with ICA-HC. $t_{\rm switch}$ denotes the frame when $\alpha$ and $\nu$ were switched.}
\renewcommand{\tabcolsep}{4.45pt}
\begin{center}
\footnotesize
\begin{tabular}{c|ccccccc}
\hline \hline
\textbf{Parameter} &$\alpha$    & $\gamma$ & $\gamma_\mathbf{n}$ & $\nu$ & $\epsilon$ &$a^{(p)}_{\mathbf{z},k}$ & $t_{\rm switch}\hspace{-1mm}$\\ \hline
\textbf{Value}   & $0.96\hspace{-0.5mm}\to\hspace{-0.5mm}0.99$ & 0.1 & 0.9  & $0\hspace{-0.5mm}\to\hspace{-0.5mm}0.99$ & $10^{-2}$ &$1$ & $100$\\
\hline \hline
\end{tabular}
\end{center}
\label{tab:HyperParam}
\end{table}

\begin{table*}
\renewcommand{\tabcolsep}{6.5pt}
\def\arraystretch{1.08}
\footnotesize
\caption{WERs (\%) on the CHiME-4 dataset for enhancement by MPDR, MLDR, Mask-MVDR, Mask-MLDR, Mask-P-MLDR, and proposed Mask-S-MLDR beamformers by batch and online processing with fixed steering vectors pre-estimated by batch SVE based on CGMMs of observations masked by all-one and NN masks. Iter. indicates whether the corresponding beamformer requires iterative updates.}

\begin{center}
\footnotesize
\begin{tabular}{c|c|c||c|cccc|c||cccc|c}
\hline \hline
\multirow{3}{*}{\textbf{Beamformer}}&\multirow{3}{*}{\textbf{Mask}} &\multirow{2}{*}{\textbf{SVE}}& \multicolumn{6}{c||}{\textbf{Batch Processing}}& \multicolumn{5}{c}{\textbf{Online Processing}} \\
\cline{4-14}
{}&{}&{\multirow{2}{*}{\textbf{(Batch)}}}&\multirow{2}{*}{\textbf{Iter.}}&\multicolumn{2}{c}{\textbf{ Dev. } }&\multicolumn{2}{c|}{\textbf{ Test } } & \multirow{2}{*}{\textbf{Avg.}}&\multicolumn{2}{c}{\textbf{ Dev. } }&\multicolumn{2}{c|}{\textbf{ Test } } & \multirow{2}{*}{\textbf{Avg.}}\\
& & & &{simu.} & {\hspace{-0mm}real\hspace{0mm}} & {simu.} & {\hspace{-0mm}real\hspace{-0mm}} & &{simu.} & {\hspace{-0mm}real\hspace{0mm}} & {simu.} & {\hspace{-0mm}real\hspace{-0mm}} & \\
\hline
MPDR   & - & \multirow{2}{*}{CGMM}& & 2.94 & {\bf 3.17} & 4.78 & 5.63 & 4.13  & 9.91 & {8.92} & \hspace{-1.2mm}12.78 & \hspace{-.3mm}14.53 & \hspace{-1.2mm}11.36\\
MLDR & - & & \checkmark & {\bf 2.72} & 3.38 & {\bf 4.39} & {\bf 5.36} & {\bf 3.96}  & {\bf 2.87} & {\bf 3.63} & {\bf 4.23} & {\bf ~5.82} & {\bf 4.14}\\
\hline
MPDR   & - &&& 2.67 & {2.53} & 3.71 & 4.29 & 3.30  & 7.63 & {6.87} & 9.80 & ~9.82 & 8.53\\
MLDR & - &\multirow{2}{*}{NN}& \checkmark & 2.42 & 2.54 & 3.18 & 3.67 & 2.95  & 2.52 & 2.73 & 3.27 & ~4.06 & 3.15\\
Mask-MVDR & NN &\multirow{2}{*}{+}&& {2.42} & {2.51} & 3.22 & {3.89} & 3.01  & {2.72} & {2.87} & 3.80 & ~{5.18} & 3.64\\
Mask-MLDR & NN &\multirow{2}{*}{CGMM}&& {\bf 2.32} & {2.52} & {3.20} & {3.65} & {2.92}  & {2.58} & {2.61} & {3.13} & ~{4.04} & {3.09}\\ 
Mask-P-MLDR & NN &&\checkmark& 2.37 & 2.48 & {\bf 2.95} & 3.60 & 2.85  & 2.49 & 2.62 & 3.07 & ~3.90 & 3.02\\
{Mask-S-MLDR} & NN &&\checkmark& {2.36} & {\bf 2.42} & 3.00 & {\bf 3.59} & {\bf 2.84}   & {\bf 2.47} & {\bf 2.56} & {\bf 2.97} & ~{\bf 3.78} & {\bf 2.95}\\

\hline \hline

\end{tabular}
\end{center}
\label{tab:WER1}
\end{table*}

Next, we compared the proposed online algorithm of joint beamforming and SVE with conventional online enhancement methods in Table \ref{tab:On_WER}. 
For the online RLS methods based on frame-by-frame processing, the forgetting factor $\alpha$ was initially set to $0.96$ and switched to $0.99$ at the 100th frame to enhance the initial convergence. For the recursive estimation of the TVVs, the smoothing factor $\gamma$ for TVV estimation was fixed to 0.1. In the online processing, the floor value $\epsilon$ for the target mask $\mathcal{M}_k(t)$ was required to replace $\mathcal{M}_k(t)$ with $\max{\left\{\mathcal{M}_k(t),\epsilon \right\}}$ for robust estimation of masked inputs $\mathcal{M}_k(t)\overline{|X\hspace{-0.2mm}_k(t)|}^2$. Here, $\epsilon$ was set to $10^{-2}$.
The experiment was conducted by using NN masks for comparison\footnote{Although the NN masks~\cite{heymann16_GEV} are not attained by online processing, the experiments using these NN masks were also conducted to compare the online beamformers with more accurate steering vectors and masks.}.

We used the Mask-S-MLDR beamformer with SVE based on the ICA-HC as the proposed online method. As a conventional online beamformer with SVE, we considered the Mask-P-MLDR with the wSCM-based SVE in~\cite{cho20}. We also evaluated the proposed Mask-S-MLDR beamformers with the wSCM-based SVE. Without the masks, the Mask-P-MLDR and Mask-S-MLDR were replaced with MLDR beamformers. As online methods wihout SVE, OIVA and Mask-P-OIVA were evaluated. We also evaluated the Mask-MVDR with SVE using NN mask alone when the target masks were provided. Similar to the derivation in Subsection \ref{ICA_RLS}, the algorithm for the online OIVA was derived based on RLS. For the OIVA, Mask-P-OIVA, and the proposed SVE based on ICA-HC, the scale factor $\nu$ in (\ref{CS_online}) was initially set to zero and switched to $0.8$ without the masks and $0.99$ with the masks at the 100th frame to enhance the stability for both the conventional wSCM-based and proposed ICA-HC-based SVE methods. For the proposed SVE based on ICA-HC, the smoothing factor $\gamma_\mathbf{n}$ for recursive estimation of smoothed noise power $P_{\mathbf{n},k}(t)$ in (\ref{power_ratio_recur}) was set to 0.9. The values of hyper-parameters for the proposed method are summarized in Table \ref{tab:HyperParam}.

In Table \ref{tab:On_WER}, all online methods except OIVA improved performance compared to the baseline without any processing. Conventional OIVA's performance was unstable but significantly improved by using masked input as a prior, providing even better performance than Mask-MVDR with SVE based on NN alone. However, Mask-P-OIVA still showed higher WERs than the statistical beamformers (MLDR and Mask-S-MLDR) with SVE combined with NN, due to inferior results in real-recorded data.
With wSCM-based SVE, the WERs of Mask-P-MLDR were similar to those of the proposed Mask-S-MLDR, like the results in Table \ref{tab:WER0}.
For beamforming with SVE methods, MLDR and the proposed Mask-S-MLDR with ICA-HC-based SVE still showed improved performance.

\begin{table}
\makegapedcells
\footnotesize
\caption{Comparison of the weighting functions for six different beamformers (MPDR, MLDR, Mask-MVDR, Mask-MLDR, Mask-P-MLDR, and Mask-S-MLDR). Moving average calculation over adjacent frames for estimating TVVs and constant factors are omitted for simplicity.
}

\renewcommand{\tabcolsep}{7pt}
\def\arraystretch{1.2}
\setcellgapes{3pt}
\begin{center}
\begin{tabular}{c|c|c}
\hline \hline
\multirow{2}{*}{\hspace{-1mm}\textbf{Beamformer}\hspace{-1mm}}&\multirow{2}{*}{{$\phi_k(\tau)$}} & \multirow{2}{*}{\textbf{\hspace{-1mm}Estimations of} $\lambda_k(\tau)$\hspace{-1mm}} \\ 
& & \\
\hline
{MPDR} & \multirow{1}{*}{const.} & \multirow{1}{*}{-} \\
\hline
MLDR& ${1}/{\lambda_k(\tau)}$&${|Y_k(\tau)|^2}$\\
\hline
{Mask-MVDR} & \multirow{1}{*}{$1-\mathcal{M}_k(\tau)$} & \multirow{1}{*}{-}\\
\hline
Mask-MLDR &${1}/{\lambda_k(\tau)}$&$\mathcal{M}_k(\tau)\overline{|X\hspace{-0.2mm}_k(\tau)|}^2$\\ 
\hline
\text{\hspace{-.5mm}Mask-P-MLDR\hspace{-.5mm}} &$\hspace{-1mm}{1}/{\lambda_k(\tau)}$&\text{\hspace{-.5mm}${{{|\hspace{-.2mm}Y\hspace{-.3mm}_k(\hspace{-.3mm}\tau)\hspace{-.2mm}|}^2 \hspace{-0.8mm} + \hspace{-0.4mm}\mathcal{M}_k(\hspace{-.3mm}\tau\hspace{-.2mm})\overline{|\hspace{-.2mm}X\hspace{-0.2mm}_k(\hspace{-.3mm}\tau)\hspace{-.2mm}|}^2}}$\hspace{-1mm}}\\ 
\hline

Mask-S-MLDR & \text{\hspace{-1mm}$\dfrac{1}{\hspace{-.3mm}\sqrt{\hspace{-.3mm}\lambda(\hspace{-.2mm}k,\hspace{-.2mm}\tau\hspace{-.2mm})}|Y\hspace{-.5mm}(\hspace{-.2mm}k,\hspace{-.4mm}\tau\hspace{-.2mm})|}$\hspace{-1mm}}& $\mathcal{M}_k(\tau)\overline{|X\hspace{-0.2mm}_k(\tau)|}^2$\\

\hline \hline
\end{tabular}
\end{center}
\label{tab:phi}
\end{table}

\begin{table*}
\def\arraystretch{1.08}
\footnotesize
\caption{WERs (\%) on the CHiME-4 dataset for enhancement by the beamformers with alternately estimated steering vectors by the proposed SVE using NN masks as the target masks to estimate TVVs and steering vectors for two different source models and four different spatial constraints. The pairs of numbers in parentheses represent the penalty weights of distortionless and null constraints. Infinite penalty weights correspond to the constraints with Lagrange multipliers.
}
\begin{center}
\begin{tabular}{c|c|c||cccc|c||cccc|c}
\hline \hline
\multirow{3}{*}{\textbf{Mask}}&\multirow{2}{*}{\hspace{-1.3mm}\textbf{Source model}\hspace{-1.3mm}}&\multirow{2}{*}{\hspace{-0.8mm}\textbf{Spatial constraints}\hspace{-0.8mm}} &\multicolumn{5}{c||}{\textbf{Batch processing}} &\multicolumn{5}{c}{\textbf{Online processing}} \\ \cline{4-13}
{}&\multirow{2}{*}{\textbf{(Beamformer)}}&\multirow{2}{*}{\textbf{$({a}_{1k}^{(p)}\hspace{-.5mm}, {a}_{\mathbf{z},k}^{(p)}\hspace{-0mm})$}}&\multicolumn{2}{c}{\textbf{ Dev. } }&\multicolumn{2}{c|}{\textbf{ Test } } & \multirow{2}{*}{\textbf{Avg.}}&\multicolumn{2}{c}{\textbf{ Dev. } }&\multicolumn{2}{c|}{\textbf{ Test } } & \multirow{2}{*}{\textbf{Avg.}}\\
&& &{simu.} & {real} & {simu.} & {\hspace{-0mm}real} & &{simu.} & {real} & {simu.} & {\hspace{-0mm}real} \\
\hline
\multirow{8}{*}{NN}&& ICA-PC (10,1) & {2.40} & {2.61} & {3.35} & {3.90} & {3.07}  & {2.93} & {3.27} & {3.84} & {4.82} & {3.72}\\ 
&\multirow{1}{*}{Gaussian Distribution}& ICA-PC (50,1) & {2.36} & {2.50} & {3.07} & {3.59} & {2.88}    & {2.89} & {3.24} & {3.81} & {4.68} & {3.66}\\ 
&\multirow{1}{*}{(Mask-P-MLDR)}& ICA-LC (\hspace{-0.3mm}$\infty\hspace{-.2mm},\hspace{-0.4mm}\infty$\hspace{-0.3mm}) & {2.44} & {2.86} & {3.29} & {4.32} & {3.23}     & {2.77} & {3.01} & {3.75} & {4.67} & {3.55}\\ 
&& ICA-HC\hspace{.6mm}($\infty$,1) & {2.37} & {2.54} & {3.07} & {3.54} & {2.88}     & {2.75} & {2.92} & {3.76} & {4.60} & {3.51}\\ 
\cline{2-13}
&& ICA-PC (10,1) & {2.48} & {2.59} & {3.62} & {3.81} & {3.13}    & {3.12} & {3.00} & {4.44} & {4.88} & {3.86}\\
&\multirow{1}{*}{Laplacian Distribution}& ICA-PC (50,1) & {2.31} & {2.47} & {3.18} & {3.50} & {2.87}    & {2.73} & {2.83} & {3.57} & {4.51} & {3.41}\\
&\multirow{1}{*}{(Mask-S-MLDR)}& ICA-LC (\hspace{-0.3mm}$\infty\hspace{-.2mm},\hspace{-0.4mm}\infty$\hspace{-0.3mm}) & {2.45} & {2.74} & {3.31} & {4.19} & {3.17}     & {2.67} & {\bf 2.82} & {3.39} & {4.41} & {3.32}\\
& & ICA-HC\hspace{.6mm}($\infty$,1) & {\bf 2.29} & {\bf 2.43} & {\bf 3.05} & {\bf 3.44} & {\bf 2.80}     & {\bf 2.67} & {2.83} & {\bf 3.38} & {\bf 4.27} & {\bf 3.29}\\

\hline
\hline

\end{tabular}
\end{center}
\label{tab:WER3}
\end{table*}

\subsection{Comparison of Various Beamformers with Fixed Steering Vectors Estimated by Batch CGMM}
\label{Exp:CGMM_BF}

Table~\ref{tab:WER1} summarizes the WERs of six different beamformers based on both batch and online processing, using fixed steering vectors pre-estimated by the batch CGMM-based SVE. In addition to the five beamformers (MPDR, MLDR, Mask-MVDR, Mask-P-MLDR, and Mask-S-MLDR) in Table~\ref{tab:WER0}, Mask-MLDR, which uses TVVs directly estimated by masked inputs of (\ref{MLDR_Mask}), was also compared. Assuming that the NN masks for target speech were available, the CGMM-based SVE was also performed for observations masked by the target masks.

Regardless of which steering vectors were given in Table \ref{tab:WER1}, the MLDR showed better performance than the MPDR by considering the non-stationarity of the target speech without masks, especially for online processing. In general, all the beamformers using steering vector estimates for observations masked by NN masks outperformed those using steering vector estimates for unmasked observations. This demonstrates that the accuracy of steering vector estimates was improved by using the masks.
Compared to Mask-MVDR, MLDR and Mask-MLDR showed improved performance, which means proper statistical modeling with TVV estimation on target speech sources can be effective enough to outperform Mask-MVDR. Furthermore, regardless of the masks used, MLDR and Mask-MLDR showed comparable performance. This means that TVV estimation of MLDR is sufficiently accurate, resulting in robust performance as long as steering vectors are estimated accurately with the masks. In particular, Mask-P-MLDR and Mask-S-MLDR utilized both masked observations and beamformer outputs, achieving robustness unlike the other beamformers that showed higher WERs.

Because the weighting function $\phi_k(\tau)$ is critical for the beamformer performance, the analysis of $\phi_k(\tau)$ is meaningful to understand the beamformers. Table \ref{tab:phi} summarizes the calculations of $\phi_k(\tau)$ for all the compared beamformers. According to (\ref{MVDR_NSCM}), the weighting function for the Mask-MVDR can be considered as $\phi_k(\tau)=1-\mathcal{M}_k(\tau)$, which is completely determined by the mask value. For the MLDR including Mask-MLDR and Mask-P-MLDR, speech-dominant t-f segments are close to zero because $\phi_k(\tau)$ is given as the reciprocal value of a TVV. Especially, $\phi_k(\tau)$ of the Mask-MLDR uses observation powers in addition to the mask value. As a result, the Mask-MLDR was able to secure robustness in estimating wSCMs derived from proper speech modeling with TVVs unlike the Mask-MVDR, which was confirmed by Table \ref{tab:WER1} where the Mask-MLDR obtained lower WERs than the Mask-MVDR. Nevertheless, the weighting function of the Mask-MLDR is calculated by masked observations without considering beamforming outputs while that of the MLDR entirely depends on the beamforming outputs. 
Note that beamforming methods are iterative if their beamformed output $Y_k(\tau)$ is utilized in the calculation of the weighting function $\phi_k(\tau)$ or the estimation of TVVs $\lambda_k(\tau)$. Otherwise, the beamforming weights are uniquely determined without iterations based on mask values. 
On the other hand, the Mask-P-MLDR and Mask-S-MLDR can achieve more robust beamforming by using wSCMs estimated from both masked observations and beamforming outputs as shown in Table \ref{tab:WER1}. Using the prior distribution for TVVs, the Mask-P-MLDR utilizes both the values in the form of addition in TVV estimation, whereas the Mask-S-MLDR does so in the form of multiplication in the calculation of the weighting function based on the assumption of source sparsity.

\subsection{Evaluation of Joint Beamforming and SVE Based on Proposed Frameworks}
\label{Exp_ICA_SVE}

In Table \ref{tab:WER3}, we evaluated joint beamforming and SVE according to the types of spatial constraints for both Gaussian and Laplacian source models when steering vector estimates were not available in advance. The experiment was conducted for both batch and online processing. When the Gaussian source models using (\ref{MLDR_pdf}) and $q\left( \mathbf{z}_k\left(\tau \right) \right)\propto \exp(-\left\|\mathbf{z}_k(\tau)\right\|_2^2)$ were used as the conventional ones, beamforming filters were obtained by the Mask-P-MLDR beamformer. On the other hand, the Laplacian source models using (\ref{ICA_pdf}) and $q\left( \mathbf{z}_k\left(\tau \right) \right)\propto \exp\left(-\left\|\mathbf{z}_k(\tau)\right\|_2\right)$ were used for the proposed Mask-S-MLDR beamformer. These beamformers were spatially constrained by three types of methods (ICA-PC, ICA-LC, and ICA-HC). 
While the penalty weights for null constraints \vspace{-.1mm} ${a}_{\mathbf{z},k}^{(p)}$ were still experimentally set to 1, the penalty weights for distortionless constraints ${a}_{1k}^{(p)}$ were varied from 10 to infinity for ICA-PC or ICA-HC. Specifically, pair values of the penalty weights of distortionless and null penalties are indicated in each row of Table \ref{tab:WER3}. Because the constraints with Lagrange multipliers correspond to the special cases of ICA-PC with infinite penalty weights, they are marked by infinite values of penalty weights ($\infty$,$\infty$) while the pair ($\infty$, 1) means the constraints based on ICA-HC. Unlike the former experiment, the recursive estimation using $P_{\mathbf{n},k}(t)$ was omitted to assess the pure estimation performance of the ICA, which was equivalent to setting $\gamma_\mathbf{n}$ to 0. The WERs were evaluated using NN masks as the target masks for estimating TVVs and steering vectors.

Regardless of the types of masks and source models, overall performance improved as the distortionless constraint was getting stricter with a weak null constraint (${a}_{\mathbf{z},k}^{(p)}\hspace{-.5mm}=1$). It might be because strong constraints enhanced the stability of beamformers by preventing undesirable distortions in beamforming outputs. 
For both the source models in batch processing, the ICA-PC with ${a}_{1k}^{(p)}\hspace{-.5mm}=50$ outperformed the ICA-LC. Too strict null constraints, in addition to strict distortionless constraints, resulted in degraded recognition performance with the ICA-LC, because strict constraints prevent the accurate SVE obtained from the demixing matrix outputs. The WERs for online processing rise because of instability and instantaneous convergence in the frame-by-frame update of beamforming weights and steering vectors compared to the batch processing using sufficiently converged outputs after many iterations. Unlike batch processing, the ICA-LC achieved superior results to the ICA-PC due to the strict constraints in the online processing where stability is much more preferred. For the online processing, it is obvious that the strict distortionless constraint played a far more critical role in avoiding undesirable distortions in beamforming outputs. In terms of the source model, the methods with the proposed Laplacian source model, where Mask-S-MLDR beamformer and MICA models are used, mostly show improved performance. It could be attributed to the effectiveness of the Laplacian source models by considering the sparsity of target speech and noises. Regardless of the types of masks and source models, the ICA-HC improved the recognition performance by avoiding too low degrees of freedom in extracting the noises with strict distortionless constraints on target outputs. Especially, the proposed Laplacian source model with the ICA-HC showed the best performances.

\subsection{Evaluation on Another Dataset with a Different ASR Model}
\label{subsection:LibriCSS}
To assess the versatility of the proposed methods as a pre-processing step for robust ASR, we also evaluated the WER scores on LibriCSS~\cite{Chen20} as an additional real-recorded dataset. The LibriCSS dataset was recorded using a 7-channel circular microphone array with a random speaker position. We used a pre-trained ASR model provided in~\cite{Chen20} and evaluated the performance in the utterance-wise fashion with batch processing. Although the dataset was originally presented for continuous speech separation, it also provides segmented utterances for the utterance-wise evaluation. Each segmented main utterance was set to target speech with the other assumed to be an interference, which enabled us to apply our beamforming methods for the main target speech. We utilized target masks estimated by the pre-trained NN model based on the conformer in~\cite{Chen21}. 

\begin{table}
\def\arraystretch{1.08}
\footnotesize
\caption{WERs (\%) of utterance-wise evaluation on the LibriCSS dataset for the baseline without any processing for input data acquired at the center microphone and enhancement by Mask-GEV, Mask-P-OIVA, Mask-MVDR, Mask-P-MLDR, and proposed Mask-S-MLDR beamformers. SVE methods based on CGMM, wSCMs, and proposed ICA-HC were performed by observations masked by NN masks. SVE based on NN alone with Mask-MVDR was also compared. OS and 0L denote no overlap with short and long inter-utterance silence, respectively.
\vspace{-2mm}
}
\renewcommand{\tabcolsep}{4.7pt}
\begin{center}
\footnotesize
\begin{tabular}{c|c|c|cccccc}
\hline \hline
\multirow{1}{*}{\hspace{-1.5mm}\textbf{Enhancement}\hspace{-1.5mm}}&\multirow{2}{*}{\hspace{-1mm}\textbf{Mask}\hspace{-1mm}}&\multirow{2}{*}{\hspace{-0.5mm}\textbf{SVE}\hspace{-0.5mm}}&\multicolumn{6}{c}{\textbf{ Overlap Ratio (\%) } }\\
\textbf{method}& & & {\hspace{-0.7mm}0S\hspace{-0.7mm}} & {0L} & {10} & {20} & {30} & {40}\hspace{-1mm}\\
\hline
\text{\hspace{-1.5mm}No Processing\hspace{-1.5mm}} & - & - & 11.8 & \hspace{-1.2mm}11.7 & \hspace{-1.2mm}18.8 & \hspace{-1.5mm}27.2 & \hspace{-1.5mm}35.6 & \hspace{-1.5mm}43.3\hspace{-1mm}\\
\hline
\text{\hspace{-1mm}Mask-GEV\hspace{-0.8mm}} &NN& - & \hspace{1.2mm}7.7 & 7.8 & 9.3 & \hspace{-1.5mm}11.4 & \hspace{-1.5mm}13.8 & \hspace{-1.5mm}15.6\hspace{-1mm}\\
\text{\hspace{-1mm}Mask-P-OIVA\hspace{-0.8mm}} &NN& - & \hspace{1.2mm}7.9 & 8.0 & 9.5 & \hspace{-1.5mm}11.7 & \hspace{-1.5mm}14.5 & \hspace{-1.5mm}16.4\hspace{-1mm}\\
\text{\hspace{-1mm}Mask-MVDR\hspace{-0.8mm}} &NN&NN& \hspace{1.2mm}{7.2} & {7.3} & 9.5 & \hspace{-1.5mm}12.0 & \hspace{-1.5mm}14.2 & \hspace{-1.5mm}16.0\hspace{-1mm}\\
\text{\hspace{-1mm}Mask-MVDR\hspace{-0.8mm}} &NN&\text{\hspace{-0.8mm}NN+CGMM\hspace{-0.8mm}}& \hspace{1.2mm}{\bf 6.1} & {\bf 6.3} & 8.1 & \hspace{-1.5mm}10.7 & \hspace{-1.5mm}12.9 & \hspace{-1.5mm}14.7\hspace{-1mm}\\
\text{\hspace{-1mm}Mask-P-MLDR\hspace{-1mm}} &NN&\text{\hspace{-0.8mm}NN+wSCM\hspace{-0.8mm}}& \hspace{1.2mm}6.8 & 6.9 & {8.2} & \hspace{-1.5mm}10.2 & \hspace{-1.5mm}13.4& \hspace{-1.5mm}14.5\hspace{-1mm}\\ 
{Mask-S-MLDR} &NN&\text{\hspace{-0.8mm}NN+CGMM\hspace{-0.8mm}}& \hspace{1.5mm}{6.1} & \hspace{0mm}{6.5} & \hspace{-0mm}{8.0} & \hspace{0mm}{9.9} & \hspace{-1.5mm}{12.7} & \hspace{-1.5mm}13.9\hspace{-1mm}\\
{Mask-S-MLDR} &NN&\text{\hspace{-0.8mm}NN+wSCM\hspace{-0.8mm}}& \hspace{1.2mm}6.9 & 7.0 & 8.1 & \hspace{-1.5mm}10.0 & \hspace{-1.5mm}12.3 & \hspace{-1.5mm}13.8\hspace{-1mm}\\ 
{Mask-S-MLDR} &NN&{\text{\hspace{-1.2mm}{\hspace{.5mm}NN+ICA-HC}\hspace{-1mm}}}& \hspace{1.2mm}{6.3} & {6.8} & {\bf 7.8} & {\bf 9.7} & \hspace{-1.5mm}{\bf 12.0} & \hspace{-1.5mm}{\bf 13.7}\hspace{-1mm}\\

\hline \hline
\end{tabular}
\end{center}
\label{tab:LibriCSS}
\end{table}

We used a 512-sample length and 256-sample shift for the STFT to adjust to the NN masks obtained by the pre-trained conformer. Similar to Table \ref{tab:WER0}, we compared Mask-GEV and Mask-P-OIVA as conventional BSE methods and Mask-P-MLDR with wSCM-based SVE and Mask-MVDR with CGMM-based SVE as conventional beamforming methods. 
As a proposed beamforming method, Mask-S-MLDR with SVE based on ICA-HC in addition to CGMM and wSCMs was evaluated.
Unlike the former experiments based on the CHiME-4 dataset, $\mathbf{W}_k$ was initialized to the identity matrix in the Mask-P-OIVA and the proposed method because the direction of the target speaker was random. All the other parameters were set identically.

In Table \ref{tab:LibriCSS}, the results are evaluated depending on the overlap ratio of interference speakers in the utterance-wise evaluation. From the results, we can observe that the WERs of input data increased with the overlap ratio. Degrees of improvement with the enhancement methods also increased with the overlap ratio by suppressing the interference. For the original data, all the enhancement methods improved the recognition performance even for no-overlap (0S, 0L) situations because the original data include inherent distortions due to distant recordings. 
In the beamforming methods, the proposed Mask-S-MLDR with SVE based on CGMM and wSCMs generally showed better performance than the conventional beamformers. Moreover, the proposed method (Mask-S-MLDR with ICA-HC) generally showed the best results among the compared methods. The additional evaluation on another dataset and ASR model confirmed the versatility and effectiveness of the proposed methods.

\subsection{Evaluation on Dynamic Target Positions}
\label{Exp:dynamic}

\begin{figure}
\centering
\includegraphics[width=.92\columnwidth]{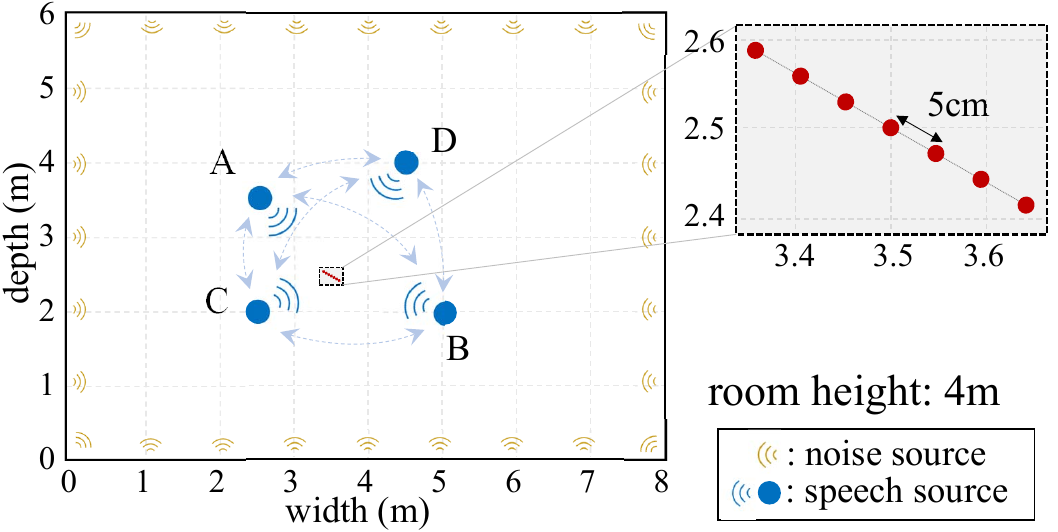}
\vspace{-1mm}
\caption{
Simulated room and microphone configuration: a target speech source moved instantaneously within an utterance between two random blue points (A, B, C, and D). A linear microphone array was simulated, with an inter-distance of 5 cm between adjacent microphones. The center microphone was fixed at (3.5 m, 2.5 m). There were 28 noise sources located at intervals of 1 m along the sides of the room. The heights of all the sources were 2 m.
}
\label{fig:rir_config}
\end{figure}

\begin{table}
\footnotesize
\caption{WERs (\%) on the simulated dataset based on Fig. \ref{fig:rir_config} for the baseline without any processing for input data acquired at the center microphone and enhancement by OIVA and MLDR without target masks and Mask-P-OIVA, Mask-MVDR, Mask-P-MLDR, and proposed Mask-S-MLDR with the masks. SVE methods based on wSCMs and proposed ICA-HC were performed by observations masked by all-one and NN masks. SVE based on NN alone with Mask-MVDR was also compared.
}
\renewcommand{\tabcolsep}{4pt}
\def\arraystretch{1.05}
\begin{center}
\footnotesize
\begin{tabular}{c|c|c||cc|cc}
\hline \hline
\multirow{2}{*}{\textbf{Beamformer}}&\multirow{2}{*}{\hspace{-.8mm}\textbf{Mask}\hspace{-.8mm}} &\multirow{2}{*}{\textbf{SVE}}& \multicolumn{2}{c|}{\hspace{-.5mm}\textbf{Moving}\hspace{-1mm}}&\multicolumn{2}{c}{\textbf{Still}} \\
 & & &{\hspace{-.5mm}Batch\hspace{-.5mm}} & {\hspace{-.5mm}Online\hspace{-.5mm}} &{\hspace{-.5mm}Batch\hspace{-1mm}} & {\hspace{-1mm}Online\hspace{-1mm}}  \\
\hline

No processing & - &-&  \multicolumn{2}{c|}{62.4} &  \multicolumn{2}{c}{62.9}\\
\hline
OIVA & - &-&  71.3& 58.0& 43.8&31.0\\
MLDR & - &wSCM&  35.8 & {\bf 27.9} & 20.1& 23.1\\
MLDR & - &ICA-HC&  36.1 & 28.8& {\bf 18.6} & 19.3\\
\hline
Mask-P-OIVA & NN & - &  23.7 & 22.2& 13.7& 16.1\\
Mask-MVDR & NN &NN & 30.3 & 36.3 & 15.6& 30.8\\
Mask-P-MLDR & NN &NN\hspace{.5mm}+\hspace{.5mm}wSCM&  18.8 & 17.4 & 12.0 & 16.6\\
Mask-S-MLDR & NN &NN\hspace{.5mm}+\hspace{.5mm}wSCM&  19.9 & 17.5 & 11.3 & 17.6\\
Mask-S-MLDR & NN &NN\hspace{.5mm}+\hspace{.5mm}ICA-HC&  20.0 & {\bf 16.9} & {\bf 10.2} & 13.3\\

\hline \hline

\end{tabular}
\end{center}
\label{tab:WER_dynamic}
\end{table}

To assess the effectiveness of online algorithms, we tested them on a simulated dynamic environment where the position of the target speaker was changed. We mixed speech sources extracted from 0S of LibriCSS and noise sources from CHiME-4 using image methods with a reverberation time of 0.2 s and a randomly chosen signal-to-noise ratio (SNR) between -5 and 0 dB. The detailed configuration is shown in Fig. \ref{fig:rir_config}. We instantaneously moved the speech position from one location to another, like~\cite{Nakashima22}. In Table \ref{tab:WER_dynamic}, we compared the batch and online enhancement methods for two scenarios: one with the position changed and another with the position fixed in an utterance.
To obtain the NN mask, we trained the conformer~\cite{Chen20}, which was originally used for continuous separation in LibriCSS. Different from Subsection \ref{subsection:LibriCSS}, we modified the conformer to output a single target mask $\mathcal{M}_k(\tau)$ as a network for speech enhancement. The conformer was trained based on a single-channel input, as in~\cite{heymann16_GEV}. All other parameters were the same as in the original conformer in~\cite{Chen20}. The network was trained using speech sources from the WSJ0 dataset~\cite{Hershey16} and noise sources from the CHiME-4 dataset. All parameters were set to the same values as in Subsection \ref{subsection:LibriCSS}, except the $\nu$ in (\ref{CS_online}) was switched to 0.8 without masks and 0.9 with masks from 0 for stable SVE in moving situations when online processing was performed.

Except for batch processing of OIVA on the moving situation, all methods improved recognition results compared to unprocessed input data. When the speech source were stationary, batch processing led to better results than online processing, except for OIVA. However, when the speech sources were moving, enhancement through online processing showed more stable performance. In this scenario, the online SVE based on wSCM showed more stable results than ICA-HC due to better convergence when target masks were not provided. When the target masks were used, Mask-P-OIVA stably enhanced the speech, achieving even better performance than the Mask-MVDR with NN-based SVE. This result was consistent with the result from Table \ref{tab:WER0}, which showed that simulation data were well enhanced by Mask-P-OIVA. The proposed ICA-HC showed comparable WERs with wSCM-based SVE when the target masks were provided. Furthermore, when the speech sources were not moving, the proposed SVE based on ICA-HC still achieved the best performance for both batch and online processing, with and without target masks.

\section{Conclusion}
\label{Conclusion}
In this paper, we presented joint beamforming and SVE methods based on ICA with distortionless and null constraints, as a pre-processing step for robust ASR. In particular, by modeling the target signal as a Laplacian distribution with TVVs, a mask-based sparse MLDR beamformer was proposed to exploit both its outputs and target masks in the weighting function of wSCMs for robust estimation. In addition, an SVE method was also derived by using the ratio of target and noise outputs of ICA optimized with the constraints. To enhance the accuracy of steering vector estimates, the strict constraints based on the Lagrange multiplier method were extended to hybrid constraints, or ICA-HC, using the power penalty as well as the Lagrange multiplier.
Moreover, an RLS-based online beamforming and SVE algorithm estimated by frame-by-frame updates was derived for practical applications. Experimental results on the various environments using CHiME-4 and LibriCSS datasets confirmed the effectiveness of the proposed methods for both batch and online processing.

\break
\appendices

\section{Update Rules of Weights in ICA-LC}
\label{appendix:ICA_LC}
$\mathbf{w}_{mk}$ to minimize (\ref{Aux_L}) can be obtained by a solution of $\nabla_{\mathbf{w}_{mk}^*} Q_k + a_{mk}^{(l)}\hspace{-.5mm}\mathbf{h}_k = 0$. In particular, $\mathbf{w}_{1k}$ for the target output is given by
\begin{eqnarray}
\mathbf{V}\hspace{-0.2mm}_k\mathbf{w}_{1k} - \mathbf{W}^{-1}_k\mathbf{e}_1 + a_{1k}^{(l)}\mathbf{h}_k = 0,
\end{eqnarray}
which can be rearranged to $\mathbf{W}\hspace{-.4mm}_k\mathbf{V}\hspace{-.4mm}_k\mathbf{w}_{1k}\hspace{-1mm}=\hspace{-1mm}(1\hspace{-.2mm}-\hspace{-.2mm}a_{1k}^{(l)})\mathbf{e}_1$.
Therefore, finding $a_{1k}^{(l)}$ gives 
$a_{1k}^{(l)} \hspace{-.5mm}=\hspace{-.5mm} 1 \hspace{-.3mm}- \hspace{-.3mm} ({\mathbf{h}_k^{H}\hspace{-.5mm}(\mathbf{W}_{\hspace{-.5mm}k}\mathbf{V}_{\hspace{-.5mm}k})^{- 1}\mathbf{e}_1}\hspace{-.3mm})^{-1},$    
and the update equation of $\mathbf{w}_{1k}$ is induced as (\ref{gen_beamform}).
Note that the constraint of $\mathbf{W}_{\hspace{-0.5mm}k}^{-1}\mathbf{e}_1=\mathbf{h}_k$ is utilized to derive (\ref{gen_beamform}). In ICA-LC of (\ref{Aux_L}), the noise outputs $\mathbf{z}_k(\tau)$ constrained to cancel the steered direction can be obtained by updating $\mathbf{w}_{mk},~2\leq m \leq M$: 
\begin{eqnarray}
\mathbf{V}_{\mathbf{z},k}\mathbf{w}_{mk} - \mathbf{W}^{-1}_k\mathbf{e}_m + a_{mk}^{(l)}\mathbf{h}_k = 0,
\end{eqnarray}
which can be rearranged to $\mathbf{W}\hspace{-0.5mm}_k\mathbf{V}\hspace{-0.5mm}_{\mathbf{z},k}\mathbf{w}_{\hspace{-0.4mm}mk} \hspace{-0.3mm} =\hspace{-0.3mm} \mathbf{e}_m \hspace{-0.3mm}-\hspace{-0.3mm} a_{mk}^{(l)}\hspace{-0.1mm}\mathbf{e}_1$.
Similar to the derivation of $\mathbf{w}_{1k}$, the Lagrange multipliers $a_{mk}^{(l)}, 2 \le m \le M,$ can be found as
\begin{equation}
a_{mk}^{(l)} = \frac{\mathbf{h}^{H}_k\left( \mathbf{W}_k\mathbf{V}_{\mathbf{z},k} \right)^{- 1}\mathbf{e}_m}{\mathbf{h}^{H}_k\left( \mathbf{W}_k\mathbf{V}_{\mathbf{z},k} \right)^{- 1}\mathbf{e}_{{1}}},
\end{equation}
which leads to the following update rules:
\begin{eqnarray}
\hspace{-10mm}\mathbf{G}_{\mathbf{z},k}^{(l)}\hspace{-2mm} &=&\hspace{-2mm}\mathbf{V}^{-1}_{\mathbf{z},k}
-\dfrac{\mathbf{V}^{-1}_{\mathbf{z},k}\mathbf{h}_k\mathbf{h}^H_k\mathbf{V}^{-1}_{\mathbf{z},k}
}{\mathbf{h}^{H}_k\mathbf{V}^{-1}_{\mathbf{z},k}\mathbf{h}_k},
\\[-1pt]
\hspace{-10mm}\tilde{\mathbf{w}}_{mk}\hspace{-2mm}& =& \hspace{-2mm}\mathbf{G}_{\mathbf{z},k}^{(l)}\mathbf{W}^{-1}_k\mathbf{e}_{{m}},
\\[3pt]
\hspace{-10mm}{\mathbf{w}}_{mk}\hspace{-2mm}&=&\hspace{-2mm}{\tilde{\mathbf{w}}_{mk}}/{\sqrt{\tilde{\mathbf{w}}_{mk}^H\mathbf{V}_{\mathbf{z},k}\tilde{\mathbf{w}}_{mk}}},~2 \leq m \leq M.
\label{noise_norm}
\end{eqnarray}

\section{Update Rules of Weights in ICA-PC and ICA-HC}
\label{derive_ICA_PC_HC}
From (\ref{cost:ICA-PC}) in ICA-PC, the optimization equation for $\mathbf{w}_{1k}$ is given as $\mathbf{H}_k\mathbf{w}_{1k} - \mathbf{W}^{-1}_{\hspace{-.5mm}k}\mathbf{e}_1 - {a}_{1k}^{(p)}\hspace{-0mm}\mathbf{h}_k = 0$,
where $\mathbf{H}_k\hspace{-.5mm}=\hspace{-.5mm}\mathbf{V}_k+{a}_{1k}^{(p)}\mathbf{h}_k\mathbf{h}^{\hspace{-.5mm}H}\hspace{-.5mm}_k$.
Then, the optimization method of $\mathbf{w}_{1k}$ can be presented as the vector coordinate descent algorithm in~\cite{Mitsui18} 
to provide the update rules:
\begin{eqnarray}
\label{VCD1}
\hspace{-14mm}\tilde{\mathbf{w}}_{1k}\hspace{-2mm}&=&\hspace{-2mm}\left(\mathbf{W}_k\mathbf{H}_k \right)^{-1}\mathbf{e}_1,
\\[2pt]
\label{VCD2}
\hspace{-14mm}{\zeta}_{1k}^{(p)} \hspace{-2mm}&=&\hspace{-2mm} \tilde{\mathbf{w}}_{1k}^H\mathbf{H}_k\tilde{\mathbf{w}}_{1k},
\\[2pt]
\label{VCD3}
\hspace{-14mm}{\psi}_k\hspace{-2mm} &=&\hspace{-2mm} {a}_1^{(p)}\hspace{-.5mm}_k\tilde{\mathbf{w}}_{1k}^H{\mathbf{h}}_k,
\\[0pt]
\label{VCD4}
\hspace{-14mm}\eta_k \hspace{-2mm}&=& \hspace{-2mm}\frac{\psi_k}{2\zeta_{1k}^{(p)}}\left(-1+{\sqrt{1+\frac{4{\zeta}_{1k}^{(p)}}{|{\psi}_k|^2}}}\right),
\\[0pt]
\label{VCD5}
\hspace{-14mm}{\mathbf{w}}_{1k}\hspace{-2mm}&=&\hspace{-2mm}\eta_k\tilde{{\mathbf{w}}}_{1k}+{a}_{1k}^{(p)}\mathbf{H}^{-1}_k \mathbf{h}_k.
\end{eqnarray}
On the other hand, the optimization equation for $\mathbf{w}_{mk}, m=2,...,M$ is expressed as
\begin{eqnarray}
\mathbf{H}_{\mathbf{z},k}\mathbf{w}_{mk} - \mathbf{W}^{-1}_k\mathbf{e}_m = 0,
\end{eqnarray}
where $\mathbf{H}_{\mathbf{z},k}\hspace{-.5mm}=\hspace{-.5mm}\mathbf{V}_{\mathbf{z},k}+{a}_{\mathbf{z},k}^{(p)}\mathbf{h}_k\mathbf{h}^H_k$.
By the conventional iterative projection algorithm in~\cite{Ono11}, $\mathbf{w}_{mk}$ can be updated by
\begin{eqnarray}
\hspace{-15mm}\tilde{\mathbf{w}}_{mk} \hspace{-2mm} &=& \hspace{-2mm}\left(\mathbf{W}_k\mathbf{H}_{\mathbf{z},k} \right)^{-1}\mathbf{e}_m,
\label{append:Noise_penalty_1}
\\[1pt]
\hspace{-15mm}{\mathbf{w}}_{mk}\hspace{-2mm}&=&\hspace{-2mm}{\tilde{\mathbf{w}}_{mk}}/{\sqrt{\tilde{\mathbf{w}}_{mk}^H\mathbf{H}_{\mathbf{z},k}\tilde{\mathbf{w}}_{mk}}},~2 \leq m \leq M.
\label{append:Noise_penalty_3}
\end{eqnarray}


Through a derivation from the extended auxiliary function for ICA-HC in (\ref{cost_prop}) similar to that in Appendix \ref{appendix:ICA_LC}, the target weights $\mathbf{w}_{1k}$ can be updated by
\begin{eqnarray}
\mathbf{w}_{1k}=\frac{\mathbf{V}^{-1}_k\mathbf{h}_k}{\mathbf{h}^H_k\mathbf{V}^{-1}_k\mathbf{h}_k} + \hat{\mathbf{w}}_{1k} - \mathbf{h}^H_k\hat{\mathbf{w}}_{1k}\frac{\mathbf{V}^{-1}_k\mathbf{h}_k}{\mathbf{h}^H_k\mathbf{V}^{-1}_k\mathbf{h}_k},
\label{w1_lp}
\end{eqnarray}
where
$\hat{\mathbf{w}}_{1k}=\left(\mathbf{W}_k\mathbf{V}_k\right)^{-1}\mathbf{e}_1$.
If $\mathbf{W}_k\mathbf{h}_k\approx \mathbf{e}_1$ by sufficiently reflecting the null constraints imposed by the power penalty, 
we have $\hat{\mathbf{w}}_{1k} \approx \mathbf{V}^{-1}_k\mathbf{h}_k$. Then, we can approximate (\ref{w1_lp}) into
\begin{eqnarray}
\mathbf{w}_{1k}\approx{\mathbf{w}}_{1k}^{(l)} = \frac{\mathbf{V}^{-1}_k\mathbf{h}_k}{\mathbf{h}^H_k\mathbf{V}^{-1}_k\mathbf{h}_k},
\label{w1_lp_simple}
\end{eqnarray}
which results in the same update equation as (\ref{gen_beamform}). On the other hand, updates of the noise weights $\mathbf{w}_{mk}, m=2, \cdots, M,$ are exactly the same as in ICA-PC.

\small
\bibliographystyle{IEEEtran}
\bibliography{IEEEabrv,Online_Mask_ICA_SVE}

\end{document}